\documentclass[sigconf]{acmart}

\usepackage{booktabs} 

\usepackage{textcomp}
\usepackage{array} 

\usepackage{float} 
\usepackage{wrapfig} 
\usepackage{flushend} 
\usepackage{cleveref}
 
\usepackage{algorithmic}

\usepackage{algorithm}

\usepackage{multirow}

\usepackage{graphicx}

\usepackage{subcaption}

\newcolumntype{P}[1]{>{\centering\arraybackslash}p{#1}}

\newcommand{\specialcell}[2][c]{%
  \begin{tabular}[#1]{@{}c@{}}#2\end{tabular}}

\setcopyright{rightsretained} 

\acmDOI{10.475/123_4}

\acmISBN{123-4567-24-567/08/06}

\acmConference[SSDBM'17]{SSDBM}{June 27-29 2017}{Chicago, Illinois,USA} 
\acmYear{2017}
\copyrightyear{2017}

\acmPrice{15.00}

\begin{document}
\title{A Unified Correlation-based Approach to Sampling Over Joins}

\author{Niranjan Kamat}
\affiliation{
  \institution{The Ohio State University}
}
\email{kamatn@cse.ohio-state.edu}

\author{Arnab Nandi}
\affiliation{
  \institution{The Ohio State University}
}
\email{arnab@cse.ohio-state.edu}

\renewcommand{\shortauthors}{N. Kamat et al.}

\begin{abstract}
Supporting sampling in the presence of joins is an important problem in data analysis, but is inherently challenging due to the need to avoid correlation between output tuples.
Current solutions provide either correlated or non-correlated samples.
Sampling might not always be feasible in the non-correlated sampling-based approaches -- the sample size or intermediate data size might be exceedingly large.
On the other hand, a correlated sample may not be representative of the join.
This paper presents a \emph{unified} strategy towards join sampling, while considering sample correlation every step of the way.
We provide two key contributions. 
First, in the case where a \emph{correlated} sample is \emph{acceptable}, we provide techniques, for all join types, to sample base relations so that their join is \emph{as random as possible}.
Second, in the case where a correlated sample is \emph{not acceptable}, we provide enhancements to the state-of-the-art algorithms to reduce their execution time and intermediate data size.
\end{abstract}
 
%
%


  

\maketitle
 
\vspace{-20pt}
\section{Introduction}

%




In cases where the data size restricts the hardware or software's ability to process it within reasonable time, sampling presents a pragmatic approach towards providing insights at scale. 
Joining multiple datasets helps incorporate related knowledge sources and develop deeper insights and has proven to be a valuable operation in data analysis in multiple scientific domains such as geo-spatial data~\cite{wickham2011asa, cho2011friendship}, sensor networks~\cite{intelberkeley}, astronomy~(SDSS~\cite{kent1994sloan}), etc.
Performing statistical analyses using aggregations is a popular step following the joins.
Sampling over joins is a compelling, yet challenging task.
Performing a join can be expensive -- sampling after materializing the join may not be a pragmatic approach.
Initial efforts in join sampling were directed towards obtaining \emph{non-correlated samples}\footnote{We define a correlated sample as one where the inclusion probability of an item is not independent from that of another~(details in Section~\ref{intro:correlated}) -- samples having any degree of randomness in the sampling process can technically be called random samples~\cite{cochran2007sampling}.}~\cite{olken1993random, acharya1999join, chaudhuri1999random}. 
As Chaudhuri et al.~\cite{chaudhuri1999random} demonstrated the inherent hardness of the problem and as online aggregation~\cite{hellerstein1997online} started gaining momentum, efforts came to be directed towards obtaining correlated, probability samples
 for aggregation queries~\cite{haas1999ripple, jermaine2008scalable, estan2006end, vengerov2015join, kandula2016quickr}.
However, non-aggregation use cases~(such as presenting a sample to user) have not been considered by this line of work. 
Correlated samples usually also have higher error estimates compared with non-correlated samples~\cite{cochran2007sampling}. 
In this context, we aim to reduce correlation in correlated samples
by maximizing \emph{join randomness}~(number of possible samples due to a join algorithm)
-- this has a side-effect of lower sampling error when compared with other correlated sampling techniques.
Further, in the case that non-correlated samples are mandated, we suggest enhancements to the state-of-the-art algorithms.
To better understand the problem domain, we illustrate the primary challenge in non-correlated sampling of joins using an example.

\vspace{-2pt}
\subsection{Challenges in Avoiding Sample Correlation}
\label{intro:motivating}
Chaudhuri et al.~\cite{chaudhuri1999random} and Gibbons et al.~\cite{acharya1999join} have provided excellent examples demonstrating one of the difficulties of join sampling -- a relation has to be sampled in a biased fashion while considering the join key cardinality of the other.
We look at another key challenge -- that of the sample size far exceeding the relation size. Before doing so, we first demonstrate the cause of sample correlation, avoidance of which results in the aforementioned challenge.


\vspace{-2pt}
\subsubsection{Correlated Samples}
\label{intro:correlated}
To avoid correlation, a tuple being present in a sample should not affect the inclusion probability of another.
For example, consider samples $S_1 = \{t_1, t_2\}$ and $S_2 = \{t_3, t_4\}$ of relations $R_1$ and $R_2$ respectively.
Join between $S_1$ and $S_2$ produces the tuples $\{t_1 \circ t_3, t_1 \circ t_4, t_2 \circ t_3, t_2 \circ t_4\}$, where $t_i \circ t_j$ represents concatenation of $t_i$ and $t_j$. 
Each tuple in $S_1$ and $S_2$ results in multiple join tuples, i.e., presence of a tuple, $t_i \circ t_j$, in the output increases the probability of other tuples having their $R_1$ component be $t_i$ or $R_2$ component be $t_j$, resulting in a correlated sample.

\vspace{-2pt}
\subsubsection{Sample Inflation}
\label{intro:inflation}
We now illustrate how avoidance of correlation can result in large sample sizes.
Consider joining two relations, each having a single distinct key, and cardinalities of $50$ and $100$, respectively.
Their join will consist of $5000$ tuples. 
A $0.1$ fraction of the join will consist of $500$ tuples -- far larger than the size of either relation. 
If a sampled tuple produces multiple joined tuples, the tuples will be correlated. 
To avoid this correlation, a sampled tuple has to be restricted to \emph{join only once}.
This results in both samples having a size of $500$, causing sampling to be counter-productive and an infeasible proposition.
We define this need to avoid correlation between output tuples resulting in increased sample size as \emph{sample inflation}. It is the main reason behind sampling over joins being difficult, and an infeasible proposition at times.

\subsubsection{Sample Inflation in Current Non-Correlated Sampling Approaches}
Chaudhuri et al.~\cite{chaudhuri1999random} provide algorithms to obtain non-correlated samples for different index and statistics availabilities.
We demonstrate how sample inflation can occur in each of these algorithms -- looking at their experimental results, it appears that they avoided sample inflation through either careful implementation or query plan optimization.
We provide enhancements to these non-correlated sampling algorithms to reduce the sample size and intermediate data size.
However, the reduction might not be satisfactory due to sample inflation.
Hence, the primary focus of this paper is to provide algorithms to reduce the correlation in correlated samples by maximizing the number of possible samples. 
This is based on our observation that non-correlated sampling techniques result in the maximum possible number of samples. Our efforts, thereby, in increasing the number of possible samples are directed towards making the samples as non-correlated as possible.
\subsection{Overview \& Contributions}


This paper provides two key contributions --  first, we look at techniques to maximize the number of possible samples in correlated sampling under the constraint of fixed sample size, when statistics over the join column are available.
We provide strategies for allocating samples to different strata of multiple relations for different join types, including equi-join, outer join, self-join, non-equi-join, and theta join for the comparators $<$, $\leq$, $>$, $\geq$~(Sections~\ref{uniformity-confidence} and \ref{uc:other}).
These techniques are derived mathematically in the Appendix. 
Although the derivations are complex, the resultant allocation strategies are simple, intuitive, and easy to compute.
They have been experimentally validated to provide allocation close to optimal allocation that was found using a brute-force search. 
The sampling error of our techniques was found to be lower than that of other correlated join sampling techniques~(Section~\ref{uc:experiment:sample-size-error}).

Our second contribution is in non-correlated sampling, where we provide enhancements to the state-of-the-art algorithms~\cite{chaudhuri1999random}.
When complete or partial statistics are available over a relation, our algorithms, \textsc{Group-Sample-Enhanced} (Section~\ref{strat:group-inflation}), and \textsc{Frequency-Partition-Sample-Enhanced}~(Section~\ref{strat:frequency-partition-inflation}),
access only half of the tuples in comparison, avoid the need to create large intermediate data which can be larger than the base relations, and remove a sampling step over the intermediate data.
In the case where statistics and indexes are available over a single relation, we specify a filter-based criterion to decide whether sampling should be used, violating which can result in the sample size exceeding the relation size~(Section~\ref{strat:stream-enhancement}). We also use this criterion to sample both relations, when statistics and indexes are available over both~(Section~\ref{sec:strat-random-sampling:final-algo}). 

\section{Related Work}
\label{current-approaches}
As performing joins can be expensive, in addition to sampling, efforts have been directed towards accelerating them through various means such as GPUs~\cite{he2008relational}, MapReduce~\cite{blanas2010comparison, zhang2012towards}, multi-threading~\cite{blanas2011design, ray2014skew}, networked execution~\cite{mokbel2004hash}, etc.
Sampling over joins has been incorporated in numerous industrial and research systems such as \emph{SQL Server}, \emph{DB2}, \emph{AQUA}~\cite{acharya1999join}, \emph{Turbo-DBO}~\cite{dobra2009turbo}, \emph{BlinkDB}~\cite{agarwal2013blinkdb}, and \emph{Quickr}~\cite{kandula2016quickr}. While these approaches target various layers of the database such as table scans, offline catalogs, and aggregation, there still exist several open opportunities. 

In the context of non-correlated sampling, Olken et al.~\cite{olken1993random} show that it is possible to commute selection with sampling, but commuting projections and joins is harder, and sample a single relation.
Chaudhuri et al.~\cite{chaudhuri1999random} provide better algorithms for joins and consider different availabilities of statistics and indexes. 
The \emph{AQUA} system~\cite{acharya1999join} obtains a simple random sample of a join in the primary key-foreign key scenario using join synopses, which \emph{TuG}~\cite{spiegel2009tug} extends to the many-to-many join scenario. While Gemulla et al.~\cite{gemulla2008linked} aim to minimize the space overhead of join synopses, \emph{CS2}~\cite{yu2013cs2} extends join synopses by proposing correlated sample synopsis, which instead of storing a sample of the join, stores a sample of the correlated tuples.
The \emph{SciBORQ} system~\cite{sidirourgos2011sciborq} maintains correlation between join attributes using impressions.
Bifocal sampling~\cite{ganguly1996bifocal} recognizes multiplicative effect of strata sizes and develops different sampling strategies based on the strata sizes to estimate  query size.


Research in obtaining non-correlated samples of joins stalled due to Chaudhuri et al.~\cite{chaudhuri1999random} showing the inherent hardness of the problem. 
However, considerable efforts have been expended towards obtaining non-correlated samples and their meaningful aggregation estimates through 
online aggregation, which allowed for error estimation during query execution \cite{haas1996hoeffding, haas1996selectivity, hellerstein1997online, haas1999ripple, qin2014pf}. 
Jermaine et al.~\cite{jermaine2005disk, jermaine2006sort} removed dependence of ripple join algorithms on the data residing in memory to estimate error.
The \emph{DBO}~\cite{jermaine2008scalable} and \emph{Turbo-DBO}~\cite{dobra2009turbo} systems allowed processing of multiple relations in a scalable fashion.
Nirkhiwale et al.~\cite{nirkhiwale2013sampling} presented the sampling algebra inherent in these techniques.
Our algorithm for sampling both sides of a join in the presence indexes, \textsc{StratJoin}, is influenced by hash ripple joins~\cite{luo2002scalable}, although they provide correlated samples.
In \emph{Wander Join}, Li et al.~\cite{li2016wander} improved ripple joins by sampling tuples from subsequent relations of the join which can join with the currently selected tuples.
In this paper, we aim to improve correlated samples by reducing correlation in samples using  strategies that maximize the number of possible samples.
Our concept of maximizing the number of samples in joins is influenced by the notion of \emph{sample randomness}, which has been introduced by Kateb et al.~\cite{al2007adaptive}, who use it to improve stratified reservoir sampling.

Some correlated sampling-based approaches such as \emph{End-Biased Sampling}~\cite{estan2006end}, \emph{Correlated Sampling}~\cite{vengerov2015join}, and \emph{Universe Sampler}~\cite{kandula2016quickr} randomly choose a range of hash of the join domain, select \emph{all} tuples whose hash of join key lies in the selected domain, and join them. 
Tuples whose hash of keys does not lie in the selected domain are discarded.
This results in the sample, and consequently the join, being a cluster sample~\cite{cochran2007sampling}, with the selected join keys defining a cluster. 
Cluster samples can be useful when join and measure columns are not correlated -- our experiments, indeed, show that such approaches have a large error in the presence of correlation between the join and measure columns.
In contrast, our approach
provides representatives from all join keys and performs well in the presence of correlation between join and measure columns.






\nocite{shin2000adaptive}
\nocite{zhao2016similarity}
\nocite{habich2005optimizing}


\vspace{-2pt}
\section{\mbox{Maximizing Randomness For Equi-Join}}
\label{uniformity-confidence}
For the case where correlated samples are acceptable, we present sample allocation techniques to maximize \emph{join randomness},
given the constraint of fixed sample size, in the presence of statistics over the join columns.
Correlated join sampling has only been studied so far in the context of aggregation queries --
this is the first work to look at it in an application-agnostic context.



\begin{table}[ht!]
\small
\centering
\caption{List of Notations}
\label{preliminaries:notation-list}
\begin{tabular}{|c | c |} 
\hline
Symbol & Explanation\\ 
\hline
\hline
$R_i$ & $i^{th}$ Relation in the Join\\ 
\hline
$N_i$ & Cardinality of $R_i$\\ 
\hline
$S_i$ & Sample of $R_i$\\ 
\hline
$n_i$ & Cardinality of $S_i$\\ 
\hline
$f$ &  Join Sampling Rate\\ 
\hline
$A$ & Join Column\\ 
\hline
$a$ & Join Value\\ 
\hline
$t.A$ & Value of Column $A$ in Tuple $t$\\
\hline
$m_i{(a)}$ & Number of Tuples in $R_i$ having Value $a$ in $A$\\  
\hline
$m_i^j$ & Number of Tuples in $R_i$ belonging to $j^{th}$ stratum\\ 
\hline
$mm_i{(a)}$ & Number of Tuples in $S_i$ having Value $a$ in $A$\\ 
\hline
$mm_i^j$& Number of Tuples in $S_i$ belonging to $j^{th}$ stratum\\ 
\hline
$z$ & Number of Relations\\ 
\hline
$n$ & Number of Strata\\ 
\hline
$k$ & Total Sample Size\\ 
\hline
$k^j$ & \specialcell{Number of Samples Allocated for Stratum $j$\\for all Relations} \\ 
\hline
\end{tabular}
\end{table}

\subsection{Join Randomness}
\label{sec:uniformity-confidence-metrics}
We use \emph{join randomness}, defined as number of samples possible in a join algorithm given the sample size, to reduce correlation in a sample.
To understand our motivation behind doing so, let us look at an example depicting the number of possible samples as a result of a few different sample allocations. 
Consider joining two relations with strata\footnote{We define \emph{strata} to be the \emph{distinct keys} of the join columns.} sizes, $mm_1^1 = 10$, $mm_2^1 = 10$, $mm_1^2 = 20$, and $mm_2^2 = 20$.
Let the sample size be constrained at 30.
The number of possible samples is given by $C^{m^{1}_{1}}_{mm^{1}_{1}} \times C^{m^{1}_{2}}_{mm^{1}_{2}} \times C^{m^{2}_{1}}_{mm^{2}_{1}} \times C^{m^{2}_{2}}_{mm^{2}_{2}}$.
Table~\ref{uc:number-of-samples-example} shows that different sample allocations can result in the number of possible samples differing by multiple orders of magnitude. 

\begin{table}[h!]
\small
\centering
\caption{\mbox{Effect of Sample Allocation on Number of Samples}}
\label{uc:number-of-samples-example}
\begin{tabular}{| c | c | c | c | c | c |} 
\hline
Index & $mm_1^1$ & $mm_2^1$ & $mm_1^2$ & $mm_2^2$ & \# Samples\\ 
\hline
\hline
$1$ & $5$ & $5$ & $10$ & $10$ & $2.2 \times 10^{15}$\\ 
$4$ & $3$ & $7$ & $8$ & $12$ & $2.3 \times 10^{14}$\\ 
$2$ & $2$ & $3$ & $12$ & $13$ & $5.3 \times 10^{13}$\\ 
$3$ & $1$ & $1$ & $14$ & $14$ & $1.5 \times 10^{11}$\\ 
\hline
\end{tabular}
\end{table}

\hspace*{12pt} If we are not restricted to sampling the input and then performing the join, but could  sample the join of the relations~(resulting in non-correlated samples), the number of possible samples would have been exceedingly large, $C^{|R_1 \bowtie R_2|}_{|S_1 \bowtie S_2|} = C^{500}_{125}$ for the first configuration.
Thus, our efforts can be perceived as making the samples as non-correlated as possible, through the metric of number of samples.
This view has been strengthened by our experiments, which show that our approach has lower error than other correlated sampling-based approaches -- non-correlated samples have theoretically lesser error than correlated samples.
\subsection{\mbox{Maximizing Randomness for Single Stratum}}
\label{sec:UC:maximizing-UC-same-strata}
Consider allocating $k^j$ tuples amongst relations $R_1$, $R_2$ ... $R_z$, each having a single stratum, to maximize the number of possible samples, $\prod_{i = 1}^{z} C^{m_i^j}_{mm_i^j}$.
Appendix~\ref{proof:single} shows that Equation~\ref{uc:formula:single} can be used for this purpose -- Section~\ref{uc:max:single} shows that it results in a low error, with the maximum difference in the sample stratum values as a result of our allocation and the optimal allocation found by searching through all possible allocations being 2. 
\begin{equation}
\label{uc:formula:single}
mm_i^j = round\left(\frac{k^j \times m_i^j}{\sum_{i = 1}^{z} m_i^j}\right)
\end{equation}



\subsection{\mbox{Maximizing Randomness for Multiple Strata}}
\label{sec:UC:maximizing-UC-multiple-strata}
We now provide the strategy to allocate a given sample size amongst different strata, in the general case of equi-join between multiple relations having multiple strata.
Consider the problem of determining the sample allocation $k^j$ for $j^{th}$ stratum, for $j \in [1, n]$, with 
$k = \sum_{j = 1}^{n} k^j$. 
Our goal is to maximize the number of possible samples,
$\prod_{j = 1}^{n} \prod_{i = 1}^{z} C^{m_i^j}_{mm_i^j}$.
Appendix~\ref{proof:multiple} shows that we can use Equation~\ref{uc:formula:multiple} to do so -- Section~\ref{uc:max:multiple} shows that the allocation is close to optimal allocation, with the maximum difference being $1$. 

\begin{equation}
\label{uc:formula:multiple}
k^j = round\left(\frac{k \times \sum_{i = 1}^{z} m_i^j}{\sum_{j = 1}^{n} \sum_{i = 1}^{z} m_i^j}\right)
\end{equation}



\subsection{Combined Algorithm -- \textsc{MaxRandJoin}}
\label{uc:maxrandjoin}
Using equations~\ref{uc:formula:single} and~\ref{uc:formula:multiple}, $mm^{j}_{i}$ can be given as follows.

\vspace{-10pt}
\begin{equation}
\label{uc:formula:combined}
mm^{j}_{i} = \frac{m^j_i}{\sum_{i = 1}^{z}m^j_i} \times \frac{k \times \sum_{i = 1}^{z}m^j_i}{\sum_{j = 1}^{n} \sum_{i = 1}^{z} m_i^j} = \frac{k \times m^j_i}{\sum_{j = 1}^{n} \sum_{i = 1}^{z} m_i^j}
\end{equation}

The resulting value will then be rounded. We can now provide our overall algorithm to maximize randomness for equi-joins.

\begin{algorithm}
\caption{Maximize Randomness for Equi-Joins}
\label{algo:mrj-equi}
\begin{algorithmic}
\STATE 1. Create $z$ \emph{allocation} tables, with each table having with $n$ rows and $3$ columns -- stratum value, the target allocation $mm^{j}_{i}$ calculated using Equation~\ref{uc:formula:combined}, and the current allocation~(at a complexity of $O(n \times z)$).
\STATE 2. Sample the $z$ relations in the following fashion. For the $i^{th}$ relation, scan through its rows to create a reservoir sample without replacement~\cite{vitter1985random} of size $mm^{j}_{i}$ for every stratum $j \in [1, n]$ using the $i^{th}$ allocation table~(at a complexity of $\sum_{i = }^{z} |R_i|$).
\STATE 3. Join the samples~(variable complexity).
\end{algorithmic}
\end{algorithm}


\vspace{-2pt}
\subsection{Applicability}
\textsc{MaxRandJoin} needs statistics over the join columns. 
If they are unavailable, table scans will be needed in the case of row-stores. However, in the case of column-stores, it will only entail a comparatively inexpensive scan over the join columns.
Complexity of joining the samples~(Step 3) is dependent on the query optimizer, which will be able to choose the correct plan since precise join column statistics will be available at this step. 

\subsection{Derivation of Allocation Strategy}
\label{uc:parameter-discussion}
We use Lagrange multipliers, a popular tool in the statistical sampling community to find approximate optimal strata allocation in closed-form under space or cost constraints~\cite{lohr2010sampling, sukhatme1957sampling, diazoptimum, kadane2005optimal, kitikidou2012optimizing}.
This includes different approaches for determining strata sizes in stratified random sampling such as Neyman allocation and cost-based allocation~\cite{lohr2010sampling}.
Lagrange multipliers provide us with critical points for maximum and minimum values, if they exist.
Our functions will have a minimum value as the number of samples is non-negative.
They will also possess a maximum value as the relation and sample sizes are bounded.
The critical points have to be plugged into the function for the number of possible samples to determine if the resultant value is a maximum or a minimum.
Our experiments~(Section~\ref{expt:maximizing-randomness}) show that suggested sample sizes 
are close to the optimal solution in practice -- with a maximum difference of $2$ in the case of single stratum allocation and $1$ for multiple strata allocation. 
The derivations possess a couple of sources of potential error.
They might result in rounding errors as they maximize the number of samples in the continuous domain, whereas the allocation occurs in the discrete domain.
Another source of error can be our use of an approximation for the harmonic sum.
Finally, we note that our \emph{simple}, intuitive, closed-form formulae maximize expressions consisting of factorials which are combinatorial in nature.






\subsection{Underlying Intuition}
While our allocation strategy for equi-joins is intuitive, its derivation is rather complex.
However, the underlying intuition behind our strategy is interesting and straightforward.
Consider allocating a fixed number of tuples randomly amongst different strata. 
Intuitively, proportional allocation is more likely to occur than other allocations.
The number of ways to come across any particular allocation will equal the number of possible samples. As proportional allocation is the most likely occurrence, proportional allocation will also result in the maximum number of possible samples.

\subsection{Comparison with Correlated Join Samplers}
\label{uc:universe-comparison}
Our approach embraces correlation in the samples, and aims to reduce it by maximizing the join randomness. 
Section~\ref{uc:experiment:sample-size-error} shows that a side-effect of maximizing the join randomness is lower sampling error when compared with other correlated samplers.



\emph{End-Biased Sampling}, \emph{Correlated Sampling}, and \emph{Universe Sampler} provide a cluster sample --
they are ill-suited for handling correlation between join and measure columns.
Such approaches circumvent the sample inflation issue by considering only those tuples that their hashing function accepts. 
This negates the need for histogram information as well.
On the other hand, we provide a unified approach to both correlated and non-correlated sampling scenarios using strategies with sound mathematical origins that tackle sampling inflation head-on.

\emph{Ripple Join}, \emph{SMS Join}, and \emph{Wander Join}, at every point in their execution, provide a correlated output as a result of a join between simple random samples of the relations.
These approaches do not take strata-based skew into consideration and as a result have a higher error than our approach~(Section~\ref{uc:experiment:sample-size-error}).

\textsc{MaxRandJoin} is specifically designed to provide a sample that is \emph{as random as possible} and takes all strata into consideration. In contrast, other correlated samplers have different goals such as responsiveness, streamability, scalability, removing the need for statistics and indexes, etc.
Comparing \textsc{MaxRandJoin} to them is not straightforward
-- they have different objectives with the resultant benefits and drawbacks. 

\section{Maximizing Randomness for Other Join Types}
\label{uc:other}
We have looked at maximizing randomness for the important case of equi-joins. 
We build upon the results and mathematical tools developed in the previous section to present techniques for maximizing randomness for all other join types.

\subsection{Outer Join \& Self-Join}
\label{uc:outer-join}
The expressions for the number of samples that need to be maximized in the case of equi-join, $\prod_{i = 1}^{z} C^{m_i^j}_{mm_i^j}$~(for single stratum) and $\prod_{j = 1}^{n} \prod_{i = 1}^{z} C^{m_i^j}_{mm_i^j}$~(for multiple strata), are clearly applicable in the case of outer joins, such as full outer join, left outer join, and right outer join, as well~(as $0! = 1$).
Therefore, the techniques used for equi-joins can be used for outer joins as well.
As self-join involves using the same relation on both sides of the join, the equi-join allocation strategy will be applicable for it as well.

\subsection{Non-Equi-Join}
\label{uc:neqjoin}
In non-equi-joins~($\neq$), a tuple can be joined with all non-matching tuples of other relations.
The number of possible samples can be given by
$\prod_{j = 1}^{n} \prod_{i = 1}^{z} \left( C^{m_i^j}_{mm_i^j} \prod_{jj = 1\ \&\ jj \neq j}^{n} \prod_{ii = 1\ \&\ ii \neq i}^{z} C^{m_{ii}^{jj}}_{mm_{ii}^{jj}} \right)$.
Interestingly, derivation in Appendix~\ref{proof:neqjoin} shows that the allocation strategy for equi-joins also works in this case.

\subsection{Theta Join}
In theta joins, tuples are joined using the provided condition over the join columns. 
We provide allocation strategies, with derivation in Appendix~\ref{proof:theta-join}, for theta joins for the common comparators,~$<$, $\leq$, $>$, and $\geq$.
Algorithm~\ref{algo:mrj-theta} provides the strategy for the $\leq$ comparator.
\begin{algorithm}
\caption{Maximize Randomness for Theta Joins ($\leq$)}
\label{algo:mrj-theta}
\begin{algorithmic}
\STATE 1. Number the strata from $1$ to $n$ in \emph{descending} order.\\
\STATE 2. Determine approximate value of $A$ using binary search, so that $k = \sum_{j = 1}^{n} \left(A^{\frac{1}{j}} \sum_{i = 1}^{z}m^j_i \right)$.\\
\STATE 3. Allocate the sample size using proportional allocation for a stratum $j$, $k^j = A^{\frac{1}{j}} \times \sum_{i = 1} ^ {z} m^j_i$.\\
\STATE 4. Allocate $k^j$ between the relations, $mm_i^j = k^j \times \frac{m^j_i}{\sum_{i = 1} ^ {z} m^j_i}$.
\end{algorithmic}
\end{algorithm}


Our derivation also shows that the only difference in the algorithm, when using the $<$ comparator instead of $\leq$ operator, is in finding $A$, such that $k = \sum_{j = 1}^{n} \left(A^{\frac{1}{j - 1}} \sum_{i = 1}^{z}m^j_i \right)$. Other steps -- 1, 3, and 4 -- remain the same.
These results can be extended for the comparators, $>$ and $\geq$, by reversing the strata order in Algorithm~\ref{algo:mrj-theta}. 

\subsection{Cross Join}
Cross join involves performing a cartesian product between the relations.
The number of possible samples can be given by $\prod_{i = 1} ^ z C^{|R_i|}_{|S_i|}$. 
This expression can be directly framed into the expression for maximizing randomness for a single stratum, $\prod_{i = 1}^{z} C^{m_i^j}_{mm_i^j}$. 
The equation for space constraint can be reframed similarly as well.
Hence, we use proportional allocation here, with
the sample size given by $|S_i| = \frac{|R_i|}{\Sigma_{j = i}^n |R_j|}$. 

\section{Non-Correlated Sampling}
\label{strat-joins-yo}
We have looked at techniques to maximize randomness of \emph{correlated} samples of joins.
Here, we look at its complementary problem -- obtaining a \emph{non-correlated} sample of a join, which is known to be a hard problem due to sample inflation~(Section~\ref{intro:inflation}).
Similar to Chaudhuri et al.~\cite{chaudhuri1999random}, we consider different availabilities of statistics and indexes. We look at some of the issues in their state-of-the-art algorithms, 
and provide enhancements to them.
We then provide an algorithm for the case of statistics and indexes being available over both relations, \textsc{StratJoin}, which minimizes the sample size.

\subsection{Enhancement to Group-Sample}
\label{strat:group-inflation}
\emph{Group-Sample} is the state-of-the-art algorithm for the case of statistics being available over one of the relations. 
We demonstrate its shortcoming~(Section~\ref{group-sample-problems}), provide an algorithm that rectifies it~(Section~\ref{group-sample-improved}), show the theoretical proof of its correctness~(Section~\ref{gsi-correctness-analysis}), provide its time~(Section~\ref{group-sample-time}) and space~(Section~\ref{group-sample-space}) complexities, and discuss a major enhancement if sorting were possible~(Section~\ref{group-sample-tradeoffs}).

\subsubsection{Issues in \emph{Group-Sample}}
\label{group-sample-problems}
We briefly describe \emph{Group-Sample} -- please refer~\cite{chaudhuri1999random} for details. Using the statistics over $R_2$, \emph{Group-Sample} samples $R_1$ in a streaming fashion, by weighting each tuple in $R_1$ by the number of tuples in $R_2$ that can join with it, resulting in $S_1$. Next, it joins $S_1$ with $R_2$, generating a group of size $m_2(t_1.A)$ for each sampled tuple $t_1$ in $S_1$. Finally, it chooses a single tuple from every group in a streaming fashion. 

Joining $S_1$ with $R_2$ will result in the intermediate materialized data having a size of $f \times \sum_{a \in Strata} m_1(a) \times m_2(a) \times m_2(a)$ on average. 
This can result in the intermediate data size exceeding the size of join between non-sampled relations -- $\sum_{a \in Strata} m_1(a) \times m_2(a)$ -- which renders sampling counter-productive.
Further, a scan is then needed over the intermediate materialized data to choose random tuples from each group, at a time complexity of $O(f \times \sum_{a \in Strata} m_1(a) \times m_2(a) \times m_2(a))$.


\subsubsection{\textsc{Group-Sample-Enhanced}}
\label{group-sample-improved}
Algorithm~\ref{algo:gsi} eliminates the need to materialize the large intermediate data and scan it.
It can be extended to the case of multiple relations, by repeating steps 2 (b) through 2 (f) for the relations having statistics.

\begin{algorithm}
\caption{\textsc{Group-Sample-Enhanced}}          
\label{algo:gsi}
\begin{algorithmic}
\STATE 1. Obtain a with-replacement sample $S_1$ of $R_1$, in a streaming fashion, by weighting a tuple $t$ by $m_2(t.A)$.
\STATE 2. While tuples of $S_1$ are streaming by do:
\STATE \hspace*{7pt} (a) Extract the next tuple, $t_1$, of $S_1$. 
\STATE \hspace*{7pt} (b) Start a scan of $R_2$. Set $i$ to $0$.
\STATE \hspace*{7pt} (c) Extract the next tuple, $t_2$, of $R_2$.
\STATE \hspace*{7pt} (d) If $t_1.A \neq t_2.A$, go to Step 2 (c). Otherwise, increment $i$.
Try to join $t_1$ with $t_2$ using \emph{Bernoulli}$\left(\frac{1}{m_2\left(t_1.A\right) - i + 1}\right)$.
\STATE \hspace*{7pt} (e) If join is successful, output the joined tuple, and go to Step 2 (a).
\STATE \hspace*{7pt} (f) If join is unsuccessful, go to Step 2 (c).
\end{algorithmic}
\end{algorithm}



\subsubsection{Proof of Correctness}
\label{gsi-correctness-analysis}
We show that the probability of $t_1$ joining with any of the $m_2(t_1.A)$ tuples equals $\frac{1}{m_2(t_1.A)}$.
Let the $i^{th}$ tuple amongst $m_2(t_1.A)$ tuples be denoted by $m_2(t_1.A)[i]$.
The probability of joining it with $t_1$ will be

\vspace*{-10pt}
\begin{flalign}
&P\left(reject\ m_2(t_1.A)[1]\right) \times P\left(reject\ m_2(t_1.A)[2]\right) \cdots & \nonumber\\& \times P\left(reject\ m_2(t_1.A)[i - 1]\right) \times P\left(accept\ m_2(t_1.A)[i]\right) \nonumber&	
\\
&= \left(1-\frac{1}{m_2(t_1.A)}\right) \times \left(1-\frac{1}{m_2(t_1.A) - 1}\right) \cdots &\nonumber\\& \times \left(1 - \frac{1}{m_2(t_1.A) + 1 - \left(i-1\right)}\right) \times \frac{1}{m_2(t_1.A) - i + 1} \nonumber&	
\\
&= \frac{m_2(t_1.A) - 1}{m_2(t_1.A)} \times \frac{m_2(t_1.A) - 2}{m_2(t_1.A) - 1} \cdots \times \frac{m_2(t_1.A) - i + 1}{m_2(t_1.A) -i + 2} &\nonumber\\& \times \frac{1}{m_2(t_1.A) - i + 1}
= \frac{1}{m_2(t_1.A)}
\nonumber&	
\end{flalign}


\subsubsection{Time Complexity} 
\label{group-sample-time}
As the initial sampling step is identical in both \emph{Group-Sample} and \emph{Group-Sample-Enhanced}, with a time complexity of $f \times \sum_{a \in Strata} m_1(a) \times m_2(a)$, we provide the average time complexity of our join step. First, we look at the probability of the join happening by the $i^{th}$ tuple.
\begin{flalign}
&P\left(join\ does\ not\ occur\ by\ m_2(t_1.A)[i]^{th}\ tuple\right) \nonumber&	
\\
& = P\left(reject\ m_2(t_1.A)[1] \right) .. \times P\left(reject\ m_2(t_1.A)[i] \right) \nonumber&	
\\
&= \left(1-\frac{1}{m_2(t_1.A)}\right) \times \left(1-\frac{1}{m_2(t_1.A) - 1}\right) \cdots &\nonumber\\& \times \left(1 - \frac{1}{m_2(t_1.A) - i + 1}\right) \nonumber&	
\\
&= \frac{m_2(t_1.A) - 1}{m_2(t_1.A)} \times \frac{m_2(t_1.A) - 2}{m_2(t_1.A) - 1} \cdots \times \frac{m_2(t_1.A) - i}{m_2(t_1.A) - i + 1} \nonumber&	
\\
&= \frac{m_2(t_1.A) - i}{m_2(t_1.A)} \nonumber&	
\\
&P\left(join\ occurs\ by\ m_2(t_1.A)[i]^{th}\ tuple\right) = &\nonumber\\& 1 - P\left(join\ does\ not\ occur\ by\ m_2(t_1.A)[i]^{th}\ tuple\right) \nonumber&	
\\
&= \frac{i}{m_2(t_1.A)} \nonumber&	
\end{flalign}

The probability of a tuple joining by the halfway stage will be $\frac{1}{2}$, as $i = \frac{m_2(t_1.A)}{2}$.
Thus, on average, half the tuples of a stratum will be accessed before the join occurs.
As a result, assuming tuples are present in a random order in $R_2$, on average, half the tuples in $R_2$ will be accessed for joining with every tuple from $S_1$.

\subsubsection{Space Complexity}
\label{group-sample-space}
Joining a tuple from $S_1$ with a random tuple from its corresponding stratum in $R_2$ does not materialize any intermediate data.
Hence, in addition to the space required for $S_1$ and the output, $f \times \sum_{a \in strata}  m_1(a) \times m_2(a)$, we  will only need to keep a count of the number of tuples belonging to the stratum that have been accessed so far~($i$ in Algorithm~\ref{algo:gsi}) in the join step.

\subsubsection{Effect of Sorted $R_2$}
\label{group-sample-tradeoffs}
In the case that $R_2$ is not sorted, the time complexity of joining $S_1$ with $R_2$ in \emph{Group-Sample} will be $O\left(f \times \left( \sum_{a \in Strata} m_1(a) \times m_2(a) \right) \times |R_2| \right)$.
While \textsc{Group-Sample-Enhanced} improves upon it, it still needs to perform a scan accessing half the tuples of $R_2$ to join with every tuple from $S_1$. 
If it were possible to sort $R_2$, the time complexities of both \emph{Group-Sample} and \textsc{Group-Sample-Enhanced} will be greatly reduced -- the number of tuples from $R_2$ that need to be accessed during the join step reducing from $O(|R_2|)$ to $O(log(|R_2|) + m_2(t_1.a))$.

\subsection{Enhancement to Frequency-Partition-Sample} 
\label{strat:frequency-partition-inflation}

\emph{Frequency-Partition-Sample} is applicable for the case where statistics are available for the larger strata of a single relation. 
It uses \emph{Group-Sample} for such strata. 
For the strata where statistics are unavailable, a join is first performed and the output is then sampled.
The space and time complexities are dominated by the larger strata.
The shortcomings of \emph{Group-Sample} will affect \emph{Frequency-Partition-Sample} as well --
the intermediate data can be expected to be large as \emph{Group-Sample} is applied over the larger strata. 
Hence, our enhancements to \emph{Group-Sample} will greatly benefit \emph{Frequency-Partition-Sample} as well.

\subsection{Enhancement to Stream-Sample}
\label{strat:stream-enhancement}

\emph{Stream-Sample} is designed for the case when one of the relations, $R_2$, has access to indexes and statistics. 
First, a with-replacement random sample, $S_1$, is constructed over $R_1$, by setting weight of a tuple $t_1$ to $m_2(t_1.A)$. Next, as the tuples of $S_1$ are streaming by, a tuple $t_1$ is joined with one of the random tuples from $m_2(t_1.A)$.

If $|S_1|$ is materialized, it will have a size $f \times |R_1 \bowtie R_2|$ on average, which can be larger than $|R_1|$, rendering sampling counter-productive. This problem occurs at a stratum level --  the sampling rate of a stratum $a$ in $R_1$ will be $f_1(a) = \frac{f \times m_1(a) \times m_2(a)}{m_1(a)} = f \times m_2(a)$. Clearly, whenever $f_1(a) > 1$, sampling will be counter-productive --
it will be prudent to not sample such strata, and only sample if it reduces the stratum size.
Note that such an approach can be used to reduce the sample size in \textsc{Group-Sample-Enhanced} as well.

\subsection{\mbox{\textsc{StratJoin} -- Sampling Both Relations}} 
\label{sec:strat-random-sampling:final-algo} 
We now provide an algorithm for the scenario where indexes and statistics are available over both relations. A simple random sample of the desired size is generated for each stratum, resulting in the output being a stratified random sample. \textsc{StratJoin} can be easily extended to multiple relations.

\begin{algorithm}
\caption{\textsc{StratJoin}}          
\label{algo:sjo}
\begin{algorithmic}
\STATE \emph{1. Sample:}  For every stratum of both relations, create a with-replacement sample of size $f \times m_1(a) \times m_2(a)$ if \mbox{$f \times m_{other}(a) \leq 1$} for a stratum $a$. Otherwise, use the entire stratum in the sample.
\STATE \emph{2. Join:} For a stratum, depending on whether none of the relations, one of the relations, or both the relations are sampled, join the samples as explained below.
\STATE \hspace*{7pt} (a) If both strata are sampled, randomly choose and join sampled tuples. Use a sampled tuple only once. 
\STATE \hspace*{7pt} (b) If a single relation is sampled, while tuples from the sampled stratum are streaming by, join them with a randomly chosen tuple from the non-sampled stratum. 
\STATE \hspace*{7pt} (c) If neither relation is sampled, join random tuples with-replacement from both strata till sampling rate is met.
\end{algorithmic}
\end{algorithm}

\section{Experimental Evaluation}
\label{experiments}
Our experiments were implemented in Java 8, and were run on an Ubuntu Linux 14.04.1 LTS system with a 24-core 2.4GHz Intel Xeon CPU, 256GB DDR3 @ 1866 MHz memory, and a 500GB @ 7200 RPM disk. 
Section~\ref{expt:maximizing-randomness} studies effectiveness of our sample allocation techniques to maximize the number of samples.
Section~\ref{expt:correlated-error} compares randomness, and error in the presence of correlation between join and measure columns, between our allocation strategies and other join techniques.

\subsection{Allocation Error}
\label{expt:maximizing-randomness}
This section looks at the effectiveness of our sample allocation techniques for maximizing the number of samples in equi-joins. 
We first validate our allocation strategy in the case of a single stratum being involved in the join and then for multiple strata. To study accuracy of our techniques, we find the best solution by searching through all possible allocations. This is computationally expensive and restricts size of datasets. We used the following metrics: mean squared error -- $\frac{\sum_{i = 1}^{n}{\left(Y_i - \hat{Y_i}\right)^2}}{n}$; mean squared relative error -- $\frac{\sum_{i = 1}^{n}{\left(\frac{Y_i - \hat{Y_i}}{Y_i}\right)^2}}{n}$; and maximum difference -- Max$\left(\left \{\left(\left|Y_i - \hat{Y_i}\right| \right): i \in [1, n]\right \}\right)$, where $Y_i$ is the actual value for optimal allocation found using brute-force search and $\hat{Y_i}$ is our predicted value.

\subsubsection{Single Stratum Partitioning}
\label{uc:max:single}

\begin{table}[h!]
\small
\centering
\caption{Error in Single Stratum Partitioning}
\label{uc:expt:single-error}
\begin{tabular}{|c | c | c | c | c |} 
\hline
Population & MSE & MSRE & Maximum Difference \\ 
\hline
\hline
150 & 0.1109 & 0.0069 & 1 \\ 
200 & 0.1197 & 0.0094 & 2 \\ 
300 & 0.1377 & 0.0135 & 2 \\ 
400 & 0.1506 & 0.0162 & 2 \\ 
500 & 0.1598 & 0.0180 & 2 \\
600 & 0.1666 & 0.0196 & 2 \\ 
700 & 0.1720 & 0.0208 & 2 \\ 
800 & 0.1761 & 0.0218 & 2 \\ 
900 & 0.1794 & 0.0226 & 2 \\ 
1000 & 0.1821 & 0.0232 & 2 \\ 
\hline
\end{tabular}
\end{table}
Since our approach might not result in the optimal allocation~(Section~\ref{uc:parameter-discussion}),
we look at its accuracy by trying different relation and strata sizes, using $3$ relations.
The total \emph{population size}, $|R_1| + |R_2| + |R_3|$, was varied from $150$ to $1000$. For each population size, all possible assignments to $|R_1|$, $|R_2|$, and $|R_3|$ were tried.
We ensured that each relation had a minimum of 5 tuples. 
The sample size was varied from $15$ to $100$. 
We found the optimal allocation through brute-force search -- searching for larger  population and sample sizes was prohibitively expensive.
Table~\ref{uc:expt:single-error} shows that all error metrics are low -- validating the effectiveness of our technique in practice for the single stratum case. 

\subsubsection{Multiple Strata Partitioning}
\label{uc:max:multiple}


We again studied the accuracy of our allocation strategy by varying the population and sample sizes.
We used 3 relations having 3 strata each, with the minimum stratum size being 3. 
In a similar fashion as above, the total population size was varied from $40$ to $55$ and the sample size was varied from $27$ to $30$ -- again, these were the upper limits for which we could find the optimal solution through brute-force search~(within a day in this case).
Table~\ref{uc:expt:multiple-error} shows that all error metrics are low. 

\begin{table}[h!]
\small
\centering
\caption{Error in Multiple Strata Partitioning}
\label{uc:expt:multiple-error}
\begin{tabular}{|c | c | c | c | c |} 
\hline
Population & MSE & MSRE & Maximum Difference \\ 
\hline
\hline
40 & 0.1439 & 0.0183 & 1 \\ 
45 & 0.1714 & 0.0221 & 1 \\ 
50 & 0.1097 & 0.0081 & 1 \\ 
55 & 0.1087 & 0.0128 & 1 \\ 
\hline
\end{tabular}
\end{table}




\subsection{Comparison with Correlated Samplers}
\label{expt:correlated-error}
We also looked at different correlated sampling techniques in the presence of correlation between join and measure columns.
Two relations with $8000$ rows were generated -- we have limited the data size to where a measurement could be obtained in around $10$ minutes. 
Each relation had 2 columns -- a join column and a measure column.
We used two common distributions, Gaussian and Zipfian, to model the column values.
We obtained similar results using the two -- results using Gaussian distribution have been presented.
The join columns were sampled from $\mathcal{N}(\mu=100, \sigma = 10)$. The measure columns were sampled from $\mathcal{N}(\mu=200, \sigma = 10)$ and $\mathcal{N}(\mu=300, \sigma = 10)$, so that we obtain non-overlapping values. 
Correlation between join and measure columns was varied from $0.1$ to $0.9$\footnote{A column having correlation $\rho$ with $X$ is generated from columns $X$ and $Y$ as $\rho \odot X \oplus (1 - \rho ^2)^{1/2} \times Y$.}.
We did not find any significant changes for differing correlation values -- results using a correlation of $0.5$ have been presented.
$50$ runs were performed, with different relations and samples being created in each run --
the median of the measurements have been presented.
We use a modification of the $L^1$-\emph{norm}, the average relative error~$\left(\sum_{i=1}^{n}\frac{|Y_i - \hat{Y_i}|}{Y_i}\right)$, to better represent the relative error.

In our context, \emph{Ripple Join}, \emph{SMS Join}, and \emph{Wander Join} have identical semantics, as do different cluster samplers such as \emph{End-Biased Sampling}, \emph{Correlated Sampling}, and \emph{Universe Sampler}.
In the figures, \emph{Stratified Random} represents stratified random sampling of the join, and gives us a non-correlated sample.
\emph{Stratified Random} suffers from sample inflation -- it provides us with the results for the best-case scenario from the perspective of correlation.

\subsubsection{Number of Possible Samples}

\begin{figure}[h!]
    \includegraphics[width=\columnwidth]{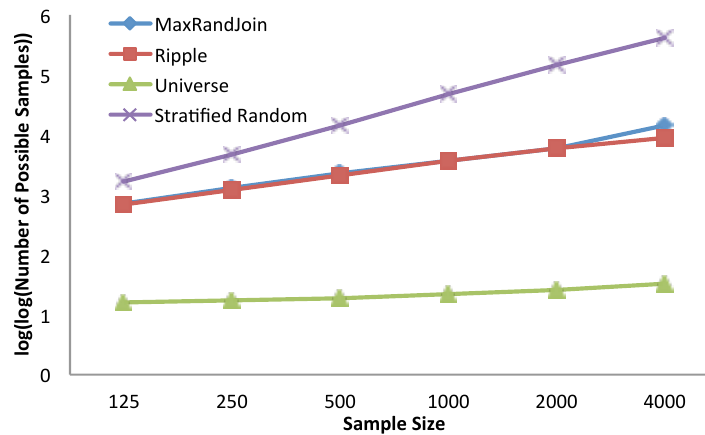}
    \caption{\emph{Stratified Random} expectedly results in the most possible samples. \textsc{MaxRandJoin} always results in more samples than \emph{Ripple Join}, while \emph{Universe Sampler} results in the fewest.}
    \label{fig:experiment:correlated:randomness}
\end{figure}
We look at the number of possible samples as a result of different techniques~(Figure~\ref{fig:experiment:correlated:randomness}), i.e. given the samples, we find $\prod_{j = 1}^{n} \prod_{i = 1}^{2} C^{m_i^j}_{mm_i^j}$. Note that the Y-axis is presented in \emph{loglog} scale~(base $e$) to better illustrate the growth pattern. 
\textsc{MaxRandJoin} consistently provides more samples than \emph{Ripple Join} and \emph{Universe}.
\textsc{StratJoin} and \emph{Stratified Random} result in the maximum possible value.




\subsubsection{Sampling Error}
\label{uc:experiment:sample-size-error}



\begin{figure}[h!]
        \includegraphics[width=\columnwidth]{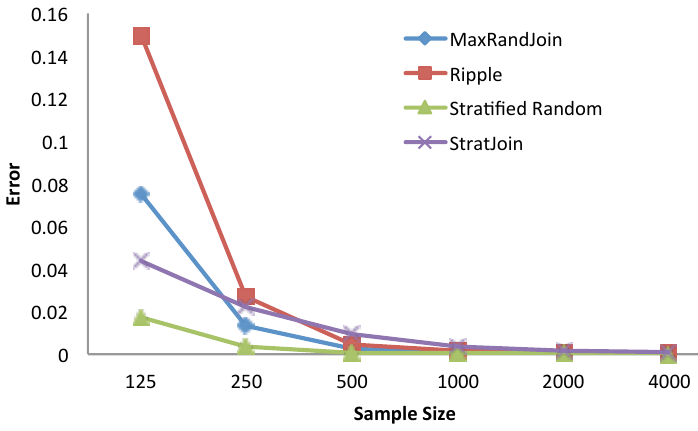}
        \caption{\textsc{MaxRandJoin} has the least error amongst correlated samplers. Interestingly, \textsc{StratJoin} has lesser error than \textsc{MaxRandJoin} for the smallest sample size.}
        \label{fig:experiment:sample-size-error}
\end{figure}

We look at the average relative error for the measure sum~(over rows) of product of the measure columns~(Figure~\ref{fig:experiment:sample-size-error}) -- other column functions that we experimented with, such as sum of measure columns and product of measure columns, gave us similar results. 
Output for \emph{Universe Sampler} has not provided as it produced large errors.
\textsc{Stratified Random} had the least error, while \textsc{MaxRandJoin} consistently had lower error than \emph{Ripple Join}.
Interestingly, \textsc{StratJoin} had lower error than \textsc{MaxRandJoin} for the smallest sample size~($125$) indicating perhaps that avoiding correlation at the expense of smaller output size, which \textsc{StratJoin} does, might be the better option at lower sampling rates.



\subsubsection{Effect of Noise on \textsc{MaxRandJoin}}
\label{expt:model-error}

\begin{figure}[h!]
    \includegraphics[width=\columnwidth]{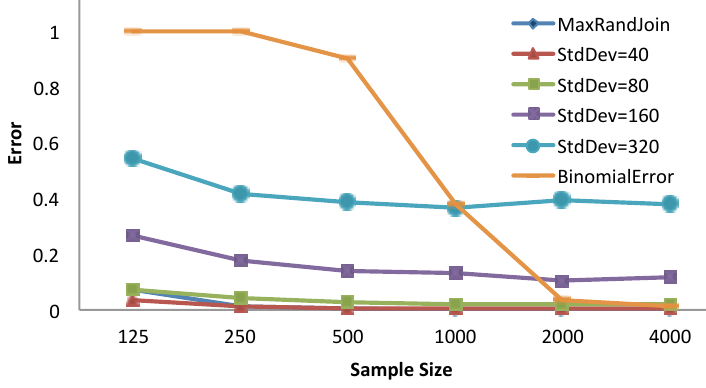}
    \caption{\textsc{MaxRandJoin} is resistant to white Gaussian noise with low standard deviation. At higher sample sizes, it can handle binomial perturbations to join count information.}
    \label{fig:experiment:noise}
\end{figure}


We investigate \textsc{MaxRandJoin}'s resistance to common types of noise -- white Gaussian and Binomial~(Figure~\ref{fig:experiment:noise}).
The standard deviation of the Gaussian noise was varied from 40 to 320.
\textsc{MaxRandJoin}'s resistance starts waning with increasing standard deviation in the Gaussian noise. It can handle Binomial error for larger sample sizes.

\subsection{Non-correlated Sampling} We have presented extensive theoretical
results describing the improvements provided by \textsc{Group-Sample-Enhanced}
over \emph{Group-Sample} in Section~\ref{strat:group-inflation}. In this
section, we demonstrate the benefits empirically, by looking at the number of
intermediate data tuples created and the time taken to obtain the join result.

\subsubsection{Experimental Setup} 
We use a similar setup to that used by Chaudhari et al.~\cite{chaudhuri1999random}.
Four tables were generated with 10000 tuples each.
The join column in each table had counts modeled using a Zipfian distribution.
The parameter $z$ of the Zipfian distributin was varied from 0 to 3.
Four other tables with 100000 tuples were generated similarly.
Each row consists of three columns -- RID (integer), JoinKey (integer), and Padding (integer). 
We have implemented the algorithms using our custom in-memory join system.
By default, we discard the first run of each experiment and report the mean of the following three runs~(the runs were nearly identical for all experiments).
In the figures, LHS refers to the relation with 10000 tuples while RHS refers to the one with 100000 tuples.
In their legends, the numbers following the algorithm name refer to the LHS and RHS skews~(when available), respectively.

\subsubsection{Intermediate Data Size}
\begin{figure}[h!]
    \includegraphics[width=\columnwidth]{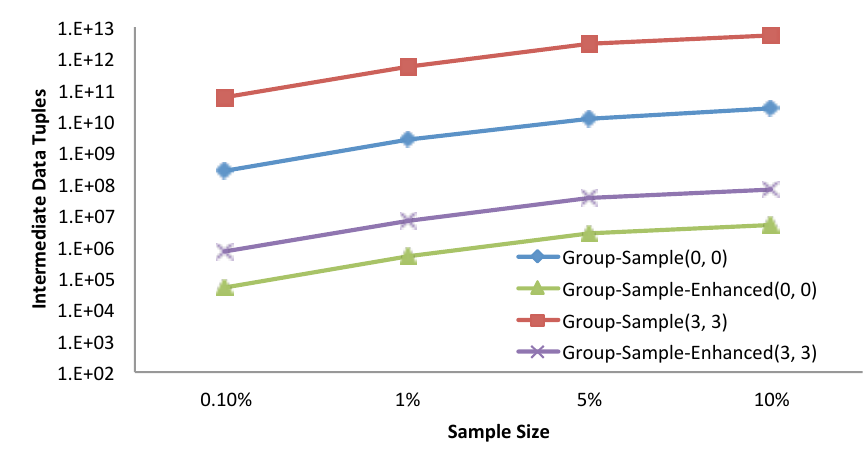}
    \caption{While an increase in the sampling rate results in a linear increase in the intermediate data size, the size required by \textsc{Group-Sample-Enhanced} is multiple orders of lesser than that of a \emph{Group-Sample} due to the usage of reservoir sampling-based techniques.} 
    \label{fig:experiment:non-correlated:intermediate:samplingRate}
\end{figure}

\begin{figure}[h!]
    \includegraphics[width=\columnwidth]{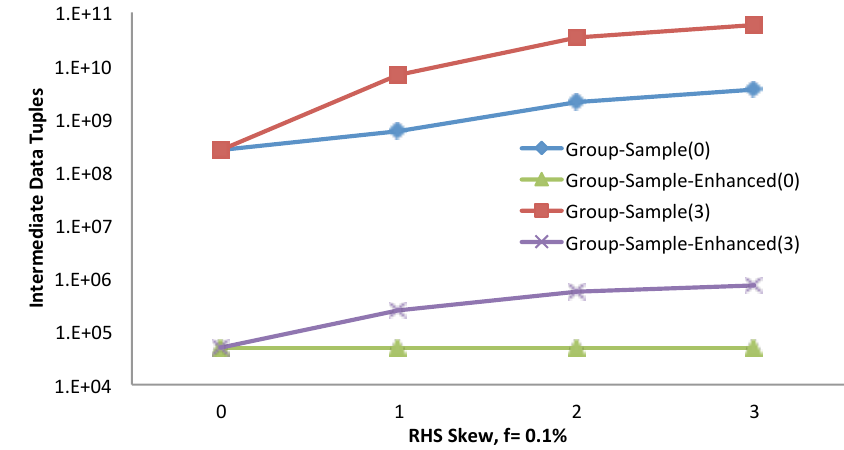}
    \caption{\textsc{Group-Sample-Enhanced} intermediate data size depends on the join size, while \emph{Group-Sample} sample size does not and can increase regardless of it~(LHS skew $ = 0$).} 
    \label{fig:experiment:non-correlated:intermediate:rhsskew}
\end{figure}

We look at the number of intermediate tuples that would be created by \emph{Group-Sample} and \textsc{Group-Sample-Enhanced} for varying sampling rates and RHS skew.
We have seen that \emph{Group-Sample} can result in large intermediate data sizes and how \textsc{Group-Sample-Enhanced} rectifies this issue -- Figures~\ref{fig:experiment:non-correlated:intermediate:samplingRate} and~\ref{fig:experiment:non-correlated:intermediate:rhsskew} provide concrete data. 
It shows that \textsc{Group-Sample-Enhanced}, whose intermediate data size is determined by the join size, can result in the intermediate data size being multiple orders of magnitude lesser when compared with \emph{Group-Sample}. Both algorithms exhibit a linear increase in the intermediate data size with  increasing sampling rate. When an increase in the RHS skew does not increase the join size~(LHS skew $ = 0$), while \textsc{Group-Sample-Enhanced} intermediate data size does no increase, that of \emph{Group-Sample} does. 

\subsubsection{Execution Time}
\begin{figure}[h!]
    \includegraphics[width=\columnwidth]{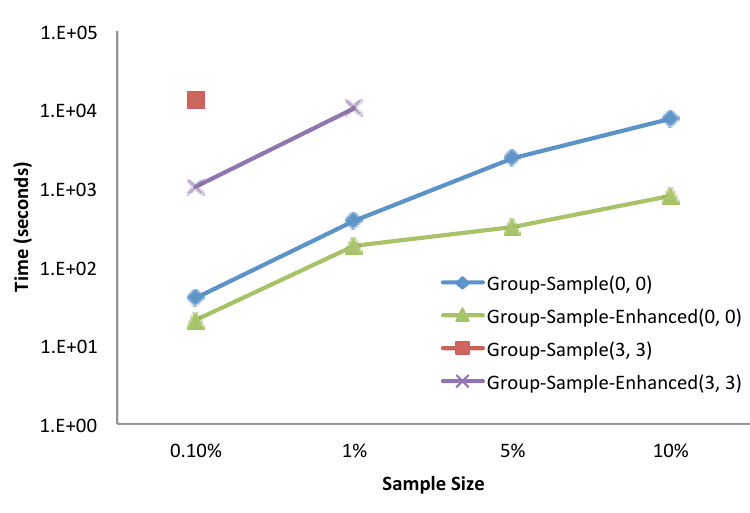}
    \caption{\textsc{Group-Sample-Enhanced} provides results around an order of magnitude sooner. It is also able to finish execution within the time limit in one of the cases where \emph{Group-Sample} is not~($f = 1\%$).} 
    \label{fig:experiment:non-correlated:time:samplingRate}
\end{figure}

\begin{figure}[h!]
    \includegraphics[width=\columnwidth]{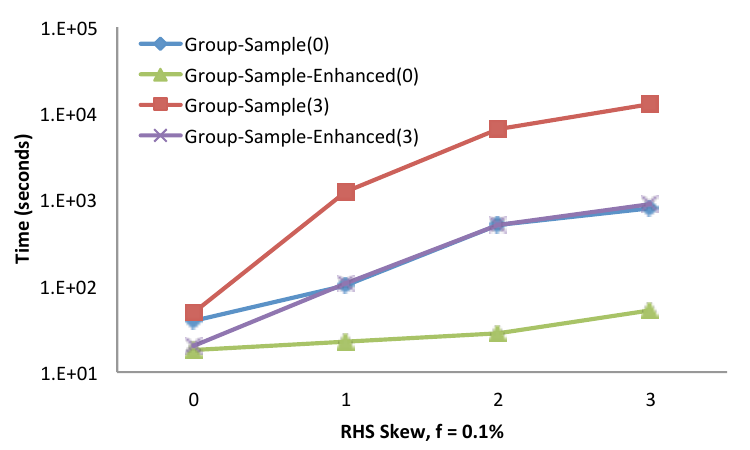}
    \caption{\textsc{Group-Sample-Enhanced} handles increasing skew better than \emph{Group-Sample}.} 
    \label{fig:experiment:non-correlated:time:rhsskew}
\end{figure}

We also compared the execution times for the two algorithms for varying sampling rate and RHS skew~(Figures~\ref{fig:experiment:non-correlated:time:samplingRate} and~\ref{fig:experiment:non-correlated:time:rhsskew}). 
The time limit for a run was capped at $4$ hours, which resulted in some of the results being unavailable.
\textsc{Group-Sample-Enhanced} usually took an order of magnitude lower time compared with \emph{Group-Sample} over changing sampling rate and skew.

\section{Conclusion \& Future Work}
\label{conclusion-future}
Join sampling is an interesting and important area of research. We have presented techniques to sample joins, in the context of correlated and non-correlated sampling -- illustrating the benefits and drawbacks in doing so. 
We provided novel techniques for increasing randomness of joins when correlated samples are acceptable. 
We showed that our techniques to maximize join randomness were effective over varying population and sample sizes.
While other correlated samplers are applicable only in the case of aggregate, approximable queries, our techniques are application-agnostic -- our techniques still result in the sample having a lower error in the aggregate measure when compared with other state-of-the-art correlated samplers.
It affirms our intuition that increasing the number of possible samples reduces the correlation in the samples.
In the context of non-correlated sampling, we provided major improvements to the state-of-the-art algorithms. We also provided an algorithm for sampling both sides of a join in the presence of statistics and indexes.

Going forward, several interesting avenues of research are enabled by our findings.
We would like to investigate the applicability of a search-based solution to obtain theoretically perfect allocation~\cite{wah1999theory}.
We would like to show a mathematical relationship between join randomness and estimate variance.
We wish that our work influences others to take correlation into consideration while designing algorithms.
Finally, we hope other research areas that use correlated samples, in both computer science and statistics communities, use strategies to maximize sample randomness.



\appendix

\section{Maximizing Randomness for Single Stratum Equi-Join}
\label{proof:single}
The notation given in Table~\ref{preliminaries:notation-list} 
 is used in the Appendix --
superscript and subscript denote the stratum index and the relation index, respectively.
Superscript has been omitted in this section, since a single stratum is considered.
The number of possible samples can be given by
\vspace{-4pt}
\begingroup
\setlength{\thinmuskip}{0mu}
\setlength{\medmuskip}{0mu}
\begin{flalign}
	&f(mm_1, \dots mm_z) = \prod_{i = 1}^{z} C^{m_i}_{mm_i} 
	= \prod_{i = 1}^{z}\frac{m_i!}{mm_i! \left(m_i - mm_i\right)!}& 
	\label{single:expanded_max}
\end{flalign}
\endgroup
The fixed sample size constraint can be given by
\begin{flalign}
\label{single:sum-constraint}
&g(mm_1, mm_2, \dots mm_z) = \sum_{i = 1}^{z} mm_i = k^1&
\end{flalign}
$g$ denotes the constraint function. 
Lagrange multipliers gives a system of $z$ equations, with $i^{th}$ ($i \in [1,z]$) equation obtained by taking partial derivative of equation \ref{single:expanded_max} with respect to $mm_i$
\begin{flalign}
\label{single:first-partial}
&f_{mm_i} = \left(\prod_{j = 1 \& j\neq i}^{z}\frac{m_j!}{mm_j! \times (m_j - mm_j)!}\right) \times m_i! \times &\nonumber \\ 
& \frac{\partial}{\partial mm_i}\left(\frac{1}{mm_i! \times \left(m_i-mm_i\right)!}\right)&
\end{flalign}
Using the basic Lagrange multipliers equation of $f_{mm_i} = \lambda \times g_{mm_i}$ and equation \ref{single:sum-constraint}, we also get
\begin{flalign}
&f_{mm_i} = \lambda \times g_{mm_i} = \lambda \times \frac{\partial \left(\sum_{i = 1}^{z} mm_i\right)}{\partial mm_i} = \lambda \label{fequalslambda}&
\end{flalign}
Using the Gamma function to represent factorial, we get
\begin{flalign}
\label{single:factorial-derivative}
&\left(n!\right)' = \Gamma'  (n+1) = n! \times \left(-\gamma + \sum_{a = 1}^{n}\frac{1}{a}\right)&	
\end{flalign}
%
where $\gamma$ is Euler-Mascheroni constant. 
$\frac{\partial}{\partial mm_i}\left(\frac{1}{mm_i! \times (m_i-mm_i)!}\right)$ can be simplified as
%
\begin{flalign}
&- \frac{ \left(mm_i! \times \frac{\partial (\left(m_i-mm_i)!\right)}{\partial mm_i}\right) + \left(\left(m_i-mm_i\right)! \times \frac{\partial (mm_i!)}{\partial mm_i}\right) }    {\left(mm_i! \times (m_i-mm_i)!\right)^2} \nonumber &\\
&= - \frac{ mm_i! \times (m_i-mm_i)! \times \left(-\gamma + \sum_{a = 1}^{m_i - mm_i}\frac{1}{a}\right)  \times (-1) }{(mm_i! \times (m_i-mm_i)!)^2} \nonumber &
\\
&- \frac{ (m_i-mm_i)! \times mm_i! \times (-\gamma + \sum_{a = 1}^{mm_i}\frac{1}{a}) }{(mm_i! \times (m_i-mm_i)!)^2} \nonumber &
\\
&= - \frac{ mm_i! \times (m_i-mm_i)! \times\left(- \sum_{a = 1}^{m_i-mm_i} \frac{1}{a} + \sum_{a = 1}^{mm_i} \frac{1}{a}\right) }{(mm_i! \times (m_i-mm_i)!)^2} \nonumber &
\\
&= - \frac{\left(\sum_{a = 1}^{mm_i} \frac{1}{a} - \sum_{a = 1}^{m_i-mm_i} \frac{1}{a}\right) }{mm_i! \times (m_i-mm_i)!}&
\end{flalign}
Plugging this into equation \ref{single:first-partial} and using equation~\ref{fequalslambda}, we get
%
\begin{flalign}
&\lambda = \left(\prod_{j = 1 \& j\neq i}^{z}\frac{m_j!}{mm_j! \times (m_j - mm_j)!}\right)  \times m_i! \times \nonumber &\\ 
&\frac{- (\sum_{a = 1}^{mm_i} \frac{1}{a} - \sum_{a = 1}^{m_i-mm_i} \frac{1}{a}) }{mm_i! \times (m_i-mm_i)!}& \nonumber \\
&= - \left(\prod_{j = 1}^{z}\frac{m_j!}{mm_j! \times (m_j - mm_j)!}\right) \left(\sum_{a = 1}^{mm_i} \frac{1}{a} - \sum_{a = 1}^{m_i - mm_i} \frac{1}{a} \right)&\nonumber 
\end{flalign}
%
The sum of harmonic series can be approximated by 
\begin{flalign}
&\sum_{a = 1}^{mm_i} \frac{1}{a} = ln(mm_i) + \gamma + \frac{1}{2 \times mm_i} - \frac{1}{12 \times mm_i^2}\nonumber  &\\
&\approx ln(mm_i) + \gamma& \label{single:harmonic-sum}
\end{flalign}
for large $mm_i$, where $\gamma$ is the Euler-Mascheroni constant.
Using the above two equations, we get 
\begingroup
\setlength{\thinmuskip}{0mu}
\setlength{\medmuskip}{0mu}
\setlength{\thickmuskip}{0mu}
\begin{flalign}
&-\lambda = \prod_{j = 1}^{z}\frac{m_j!}{mm_j! \times (m_j - mm_j)!} \times \left(ln(mm_i)  - ln(m_i - mm_i)\right) \nonumber & 
\\
&= \prod_{j = 1}^{z}\frac{m_j!}{mm_j! \times (m_j - mm_j)!} \times ln\left(\frac{mm_i}{m_i - mm_i}\right)& \label{single:using-harmonic-approx}
\end{flalign}
\endgroup
%
Note that the term $\left(\prod_{j = 1}^{z}\frac{m_j!}{mm_j! \times (m_j - mm_j)!}\right)$ is present in all $i \in [1,z]$ equations -- we denote it by $C$ and get
%
\begin{flalign}
&\frac{- \lambda} {C} = ln\left(\frac{mm_i}{m_i - mm_i}\right)&
\end{flalign}
%
Consider the equation for another relation $p$ 
\begin{flalign}
&\frac{- \lambda} {C} = ln\left(\frac{mm_p}{m_p - mm_p}\right)&
\end{flalign}
\noindent From the above two equations, we get
\begin{flalign}
&\frac{mm_i}{m_i - mm_i} = \frac{mm_p}{m_p - mm_p} = A &
\end{flalign}
\noindent for a constant A. Therefore,
\begin{flalign}
\label{single:mmi-in-terms-of-K}
&mm_i = \frac{A \times m_i}{1 + A}&
\end{flalign}
\noindent Using equations \ref{single:sum-constraint} and \ref{single:mmi-in-terms-of-K}, we get
\begin{flalign}
&k^1 = \sum_{i = 1}^{z} mm_i = \sum_{i=1}^{z} \frac{A \times m_i}{1 + A}
= \frac{A}{1+A}\sum_{i=1}^{z}m_i \nonumber  &
\end{flalign}
\noindent Using above two equations, we get 
\begin{flalign}
\label{single:final}
&mm_i = \left(\frac{A}{1+A}\right)\times m_i
=\frac{k^1 \times m_i}{\sum_{a = 1}^{z} m_a}&
\end{flalign}

\section{Maximizing Randomness for Multiple Strata Equi-Join}
\label{proof:multiple}
The constraint of fixed sample size for equi-joins with multiple strata can be given as follows, given $k^j = \sum_{i = 1}^{z} mm_i^j$.
\begin{flalign}
\label{multiple:sum-constraint}
&k = \sum_{j = 1}^{n} k^j = g\left(k^1, k^2 \dots k^n\right)&
\end{flalign}
%
%
The number of possible samples can be given by 
%
\begin{flalign}
&f\left(k^1, k^2 \dots k^n\right) = \prod_{j = 1}^{n} \prod_{i = 1}^{z} C^{m_i^j}_{mm_i^j} \nonumber & \\
&= \prod_{j = 1}^{n} \prod_{i = 1}^{z}\frac{m_i^j!}{mm_i^j! \times \left(m_i^j - mm_i^j\right)!} \nonumber  & \\
&= \prod_{j = 1}^{n}  \prod_{i = 1}^{z-1}\frac{m_i^j!}{mm_i^j! \times (m_i^j - mm_i^j)!} \times \frac{m_z^j!}{mm_z^j! \times (m_z^j - mm_z^j)!}  \nonumber &\\
&= \prod_{j = 1}^{n}  \prod_{i = 1}^{z-1}\frac{m_i^j!}{mm_i^j! \times (m_i^j - mm_i^j)!} \times \nonumber & \\ 
&\frac{m_z^j!}{(k^j - \sum_{i=1}^{z-1}mm_i^j)! \times (m_z^j - (k^j - \sum_{i=1}^{z-1}mm_i^j))!}& \label{multiple:expanded-product}
\end{flalign}
%
Equation \ref{single:final} provides optimal allocation amongst relations given allocation to a stratum, and using it in the above equation results in the number of samples increasing or staying the same.
Hence, $k^j$ $\left(j \in [1,n]\right)$ are the variables here. 
%
Using Lagrange multipliers, we get the $j^{th}$ equation as
%
\begin{flalign}
&f_{k_j} = \prod_{j = 1}^{n}  \prod_{i = 1}^{z-1}\frac{m_i^j!}{mm_i^j! \times \left(m_i^j - mm_i^j\right)!} \times m_z^j! \times \nonumber& \\ 
&\frac{\partial}{\partial k^j}\left(\frac{1}{\left(k^j - \sum_{i=1}^{z-1}mm_i^j\right)! \times \left(m_z^j - \left(k^j - \sum_{i=1}^{z-1}mm_i^j\right)\right)!}\right)& \label{original:diffeerntial}
\end{flalign}
%
We simplify $\frac{\partial}{\partial k^j}\left(\frac{1}{(k^j - \sum_{i=1}^{z-1}mm_i^j)! \times (m_z^j - (k^j - \sum_{i=1}^{z-1}mm_i^j))!}\right)$ as,
%
\begin{flalign}
&\frac{-1}{{\left(\left(k^j - \sum_{i=1}^{z-1}mm_i^j\right)! \times \left(m_z^j - \left(k^j - \sum_{i=1}^{z-1}mm_i^j\right)\right)!\right)^2}}\times \nonumber&
\\
&\left(k^j - \sum_{i=1}^{z-1}mm_i^j\right)! \times \left(-\gamma + \sum_{l = 1}^{k^j - \sum_{i=1}^{z-1}mm_i^j}\frac{1}{a}\right) \times &\nonumber\\
&\left(m_z^j - \left(k^j - \sum_{i=1}^{z-1}mm_i^j\right)\right)! - \left(k^j - \sum_{i=1}^{z-1}mm_i^j\right)! \times &\nonumber\\
&\left(m_z^j - \left(k^j - \sum_{i=1}^{z-1}mm_i^j\right)\right)! \times \left(-\gamma + \sum_{l = 1}^{m^j_z - \left(k^j - \sum_{i=1}^{z-1}mm_i^j\right)}\frac{1}{a}\right) &\nonumber
\\
&= \frac{-1 \times \left(k^j - \sum_{i=1}^{z-1}mm_i^j\right)! \times \left(m_z^j - \left(k^j - \sum_{i=1}^{z-1}mm_i^j\right)\right)!  }
{\left(\left(k^j - \sum_{i=1}^{z-1}mm_i^j\right)! \times \left(m_z^j - \left(k^j - \sum_{i=1}^{z-1}mm_i^j\right)\right)!\right)^2} \times & \nonumber\\
& \left(-\gamma + \sum_{l = 1}^{k^j - \sum_{i=1}^{z-1}mm_i^j}\frac{1}{a} + \gamma - \sum_{l = 1}^{m^j_z - \left(k^j - \sum_{i=1}^{z-1}mm_i^j\right)}\frac{1}{a}\right) & \nonumber 
\\
&= \frac{ln\left(k^j - \sum_{i=1}^{z-1}mm_i^j\right) + \gamma - ln\left(m^j_z - \left(k^j - \sum_{i=1}^{z-1}mm_i^j \right)\right) - \gamma}
{- \left(k^j - \sum_{i=1}^{z-1}mm_i^j \right)! \times \left(m_z^j - \left(k^j - \sum_{i=1}^{z-1}mm_i^j \right)\right)!} & \nonumber
\end{flalign}
\begin{flalign}
&= \frac{ln\left(\frac{m^j_z - \left(k^j - \sum_{i=1}^{z-1}mm_i^j \right)}{k^j - \sum_{i=1}^{z-1}mm_i^j} \right)}
{mm_z^j! \times \left(m^j_z - mm_z^j \right)!} &
\end{flalign}

%
Substituting above equation in equation \ref{original:diffeerntial}, we get
%
\begingroup
\setlength{\thinmuskip}{0mu}
\setlength{\medmuskip}{0mu}
\setlength{\thickmuskip}{0mu}
\begin{flalign}
&f_{k_j} = \prod_{j = 1}^{n} \prod_{i = 1}^{z}\frac{m_i^j!}{mm_i^j! \times \left(m_i^j - mm_i^j\right)!} ln\left(\frac{m_z^j - k^j + \sum_{i = 1}^{z - 1} mm_i^j}{k^j - \sum_{i = 1}^{z - 1} mm_i^j}\right) \nonumber&
\end{flalign}
\endgroup
%
From the standard Lagrange multipliers equation of $f_{k_j} = \lambda \times g_{k_j}$ and equation \ref{multiple:sum-constraint}, we get
%
\begin{flalign}
&f_{k_j} = \lambda \times g_{k_j} = \lambda \times \frac{\partial \left(\sum_{j = 1}^{n} k^j \right)}{\partial k_j}  = \lambda	&
\end{flalign}
%
From the above two equations, we get
%
\begingroup
\setlength{\thinmuskip}{0mu}
\setlength{\medmuskip}{0mu}
\setlength{\thickmuskip}{0mu}
\begin{flalign}
&\lambda = \left(\prod_{j = 1}^{n} \prod_{i = 1}^{z}\frac{m_i^j!}{mm_i^j! \times \left(m_i^j - mm_i^j \right)!}\right) ln \left(\frac{m_z^j}{k^j - \sum_{i = 1}^{z - 1} mm_i^j} - 1 \right) \nonumber&
\end{flalign}
\endgroup
%
Consider the equation for another stratum $p$
%
\begingroup
\setlength{\thinmuskip}{0mu}
\setlength{\medmuskip}{0mu}
\setlength{\thickmuskip}{0mu}
\begin{flalign}
&\lambda = \left(\prod_{j = 1}^{n} \prod_{i = 1}^{z}\frac{m_i^j!}{mm_i^j! \times \left(m_i^j - mm_i^j \right)!}\right) ln\left(\frac{m_z^p}{k^p - \sum_{i = 1}^{z - 1} mm_i^p} - 1\right) \nonumber&
\end{flalign}
\endgroup
%
From the above 2 equations, given a constant A, we get
%
\begin{flalign}
&\frac{k^j - \sum_{i = 1}^{z - 1} mm_i^j}{m_z^j} = \frac{k^p - \sum_{i = 1}^{z - 1} mm_i^p}{m_z^p} = A&
\end{flalign}
%
Using equation \ref{single:final} and the above equation, we get
%
\begin{flalign}
&A = \frac{k^j - \sum_{i=1}^{z-1}\frac{k^j \times m_i^j}{\sum_{a=1}^{z}m_a^j}}{m_z^j}
= \frac{k^j \times \left(1 - \sum_{i=1}^{z-1} \frac{m_i^j}{\sum_{a=1}^{z}m_a^j} \right)}{m_z^j}&\nonumber\\
&= \frac{k^j \times \left(\sum_{a=1}^{z}m_a^j - \sum_{i=1}^{z-1} m_i^j \right)}{m_z^j \times \sum_{a=1}^{z}m_a^j}
=\frac{k^j}{\sum_{a=1}^{z}m_a^j}&
\end{flalign}
%
%
\begin{flalign}
\label{multiple:kj-in-A}
&k^j = A \times \sum_{a = 1}^{z} m_a^j&
\end{flalign}
Using above equation and equation \ref{multiple:sum-constraint}, we get
\begin{flalign}
&k = \sum_{j=1}^{n}\left(A \times \sum_{a=1}^{z}m_{a}^{j}\right)&
\end{flalign}
\begin{flalign}
\label{A-in-k}
&A = \frac{k}{\sum_{j = 1}^{n} \sum_{a = 1}^{z} m_a^j}&
\end{flalign}
%
Using equations \ref{multiple:kj-in-A} and \ref{A-in-k}, we get
\begin{flalign}
&k^j = \frac{k \times \sum_{i = 1}^{z} m_i^j}{\sum_{j = 1}^{n} \sum_{i = 1}^{z} m_i^j}&
\end{flalign}
%


\section{Maximizing Randomness For Non-Equi-Joins}
\label{proof:neqjoin}
Number of samples for non-equi-joins can be given by
\begin{flalign}
&
\prod_{j = 1}^{n} \prod_{i = 1}^{z} \left( C^{m_i^j}_{mm_i^j} \prod_{jj = 1\ \&\ jj \neq j}^{n} \prod_{ii = 1\ \&\ ii \neq i}^{z} C^{m_{ii}^{jj}}_{mm_{ii}^{jj}} \right)
\nonumber&
\\
&
= \prod_{j = 1}^{n} \prod_{i = 1}^{z} \left( C^{m_i^j}_{mm_i^j} 
\frac{C^{m_i^j}_{mm_i^j} \prod_{j = 1}^{n} \prod_{i = 1}^{z} C^{m_i^j}_{mm_i^j} }
{\prod_{a = 1}^{n}C^{m^a_i}_{mm^a_i} \prod_{b = 1}^{z}C^{m_b^j}_{mm_b^j}}  \right)
\nonumber&
\\
&
= \prod_{j = 1}^{n} \prod_{i = 1}^{z} \left( C^{m_i^j}_{mm_i^j} 
\frac{C^{m_i^j}_{mm_i^j} K }
{\prod_{a = 1}^{n}C^{m^a_i}_{mm^a_i} \prod_{b = 1}^{z}C^{m_b^j}_{mm_b^j}}  \right)
\nonumber&
\\
&
= K^{nz} \prod_{j = 1}^{n} \prod_{i = 1}^{z} \left( C^{m_i^j}_{mm_i^j} 
\frac{C^{m_i^j}_{mm_i^j}}
{\prod_{a = 1}^{n}C^{m^a_i}_{mm^a_i} \prod_{b = 1}^{z}C^{m_b^j}_{mm_b^j}}  \right)
\nonumber&
\\
&
\cdots \textrm{where}\ K = \prod_{j = 1}^{n} \prod_{i = 1}^{z}  C^{m_i^j}_{mm_i^j}  \nonumber&
\\
&
= K^{nz} \prod_{j = 1}^{n} \prod_{i = 1}^{z}  C^{m_i^j}_{mm_i^j} 
C^{m_i^j}_{mm_i^j}
   \prod_{j = 1}^{n} \prod_{i = 1}^{z} \frac{1}{\prod_{a = 1}^{n}C^{m^a_i}_{mm^a_i} \prod_{b = 1}^{z}C^{m_b^j}_{mm_b^j}}
\nonumber&
\\
&
= K^{nz} K^2 \frac{1}{\prod_{j = 1}^{n} \prod_{i = 1}^{z} \prod_{a = 1}^{n}C^{m^a_i}_{mm^a_i}} \frac{1}{\prod_{j = 1}^{n} \prod_{i = 1}^{z} \prod_{b = 1}^{z}C^{m_b^j}_{mm_b^j} }
\nonumber&
\\
&
= K^{nz} K^2 \frac{1}{\prod_{j = 1}^{n} K} \frac{1}{\prod_{i = 1}^{z}K} 
= K^{nz} K^2 \frac{1}{K^n} \frac{1}{K^z} = K ^{nz + 2 - n - z}
\nonumber&
\end{flalign}
$K$ can be maximized using proportional allocation~(Appendix~\ref{proof:multiple}). Hence, we  use the equi-join allocation strategy here as well.

\section{\mbox{Maximizing Randomness For Theta Joins}}
\label{proof:theta-join}
For $\leq$ comparator, number of samples can be given by 
\\
$\prod_{i = 1} ^ z \prod_{j = 1} ^ {n} C^{m^j_i}_{mm^j_i} \prod_{ii = 1\ \& \ ii \neq i} ^ z \prod_{jj = 1}^{j} C^{m^{jj}_{ii}}_{mm^{jj}_{ii}}$ and its log by
\begin{flalign}
&
\sum_{i = 1} ^ z \sum_{j = 1} ^ {n} \left( ln\left(C^{m^j_i}_{mm^j_i}\right) + \sum_{ii = 1\ \& \ ii \neq i}^ z \sum_{jj = 1}^{j} ln\left(C^{m^{jj}_{ii}}_{mm^{jj}_{ii}}\right) \right)
\nonumber&
\\
&
= \sum_{i = 1} ^ z \sum_{j = 1} ^ {n}  ln\left(C^{m^{j}_{i}}_{mm^{j}_{i}}\right) + \sum_{i = 1} ^ z \sum_{j = 1} ^ {n} \sum_{ii = 1\ \& \ ii \neq i}^ z \sum_{jj = 1}^{j} ln\left(C^{m^{jj}_{ii}}_{mm^{jj}_{ii}}\right)
\nonumber&\\
&
= \sum_{i = 1} ^ z \sum_{j = 1} ^ {n}  ln\left(C^{m^{j}_{i}}_{mm^{j}_{i}}\right) + \sum_{i = 1} ^ z \sum_{j = 1} ^ {n} \sum_{ii = 1}^ z \sum_{jj = 1}^{j} ln\left(C^{m^{jj}_{ii}}_{mm^{jj}_{ii}}\right) - 
\sum_{i = 1} ^ z \sum_{j = 1} ^ {n} \sum_{jj = 1}^{j} ln\left(C^{m^{jj}_{i}}_{mm^{jj}_{i}}\right)
\nonumber&
\\
&
= \sum_{i = 1} ^ z \sum_{j = 1} ^ {n}  ln\left(C^{m^{j}_{i}}_{mm^{j}_{i}}\right) + (z - 1) \times \sum_{i = 1}^z \sum_{j = 1}^n \sum_{jj = 1} ^ j ln\left(C^{m^{jj}_{i}}_{mm^{jj}_{i}}\right)
\nonumber&
\end{flalign}
\begin{flalign}
&
= \sum_{i = 1} ^ z \sum_{j = 1} ^ {n}  ln\left(C^{m^{j}_{i}}_{mm^{j}_{i}}\right) + (z - 1) \times \sum_{i = 1}^z \sum_{j = 1}^n (n - j + 1) ln\left(C^{m^{j}_{i}}_{mm^{j}_{i}}\right)
\nonumber&
\\
&
= (z - 1) \times \sum_{i = 1}^z \sum_{j = 1}^n (n - j + 1 + \frac{1}{z - 1}) \times ln\left(C^{m^{j}_{i}}_{mm^{j}_{i}}\right)
\nonumber&
\\
&
\approx (z - 1) \times \sum_{i = 1}^z \sum_{j = 1}^n (n - j + 1) \times ln\left(C^{m^{j}_{i}}_{mm^{j}_{i}}\right)
\nonumber&
\\
&
= (z - 1) \times \sum_{i = 1}^z \sum_{j = 1}^n j \times ln\left(C^{m^{j}_{i}}_{mm^{j}_{i}}\right)
\nonumber&
\end{flalign}
Strata order was reversed in the last step.
We now have to minimize
$ \sum_{i = 1}^z \sum_{j = 1}^n j \times (ln(mm^j_i!) + ln((m^j_i - mm^j_i)!))$, with $k = \sum_{i = 1}^z \sum_{j = 1}^n mm^j_i = g(mm_1^1 ... mm^j_i)$. Lagrange multipliers gives us the $\{i, j\}^{th}$ $(i \in [1, z], j \in [1, n])$ equation 
\begin{flalign}
&
f_{mm^j_i} = \lambda \frac{\partial g}{\partial mm^j_i} = \lambda  
= \frac{\partial}{\partial mm^j_i}(j (ln(mm^j_i!) + ln((m^j_i - mm^j_i)!)))
\nonumber&\\
& = j \left( \frac{1}{mm^j_i!} mm^j_i! \left(-\gamma + \sum_{a = 1}^{mm^j_i} \frac{1}{a} \right) \right) + \nonumber&\\ 
& j\left (\frac{1}{(m^j_i - mm^j_i)!} (m^j_i - mm^j_i)! \left(-\gamma + \sum_{a = 1}^{m^j_i - mm^j_i} \frac{1}{a} \right)(-1) \right) \nonumber&\\
& = j \left(\sum_{a = 1}^{mm^j_i} \frac{1}{a} - \sum_{a = 1}^{m^j_i - mm^j_i} \frac{1}{a} \right)
\nonumber&\\
& = j (ln(mm^j_i) + \gamma - ln(m^j_i - mm^j_i) - \gamma)
= j \left(ln\left( \frac{mm^j_i}{m^j_i - mm^j_i} \right) \right)
\nonumber&\\
& = -j \left( ln \left( \frac{m^j_i}{mm^j_i} - 1\right) \right) 
\approx -j \left( ln\left( \frac{m^j_i}{mm^j_i} \right) \right)
\nonumber&
\end{flalign}

Note that the benefits of maximizing randomness diminish with increasing sampling rates.
Thus, for every equation, $\left(\frac{m^j_i}{mm^j_i}\right)^j$ is constant. 
Hence, for any stratum as $j$, the exponent, is constant, $m^j_i/mm^j_i$ will be constant -- showing that proportional allocation should be used to allocate samples for a stratum between the relations. 
Therefore, $mm^j_i = \frac{k^j m^j_i}{\sum_{i = 1}^z{m^j_i}}$. Therefore, for all strata, using above two equations, $\left(\frac{k^j}{\sum_{i = 1}^z m^j_i}\right)^j$ is constant, denoted by A. Therefore, $k^j = A^{\frac{1}{j}}\sum_{i = 1}^z m^j_i$ and $k = \sum_{j = 1}^n k^j = \sum_{j = 1}^n A^{\frac{1}{j}}\sum_{i = 1}^z m^j_i$. 
$A$ is the only unknown and can be found using binary search within error. Finding $A$ lets us find $k^j$, which leads us to finding $mm_i^j$.

\noindent \emph{$<$ Comparator:} Following similar line of reasoning, the only difference in the algorithm is in the formula for A, given by $k^j = A^{\frac{1}{j - 1}}\sum_{i = 1}^z m^j_i$. 

 
\bibliographystyle{ACM-Reference-Format}
\bibliography{document} 


\begin{thebibliography}{00}


\ifx \showCODEN    \undefined \def \showCODEN     #1{\unskip}     \fi
\ifx \showDOI      \undefined \def \showDOI       #1{{\tt DOI:}\penalty0{#1}\ }
  \fi
\ifx \showISBNx    \undefined \def \showISBNx     #1{\unskip}     \fi
\ifx \showISBNxiii \undefined \def \showISBNxiii  #1{\unskip}     \fi
\ifx \showISSN     \undefined \def \showISSN      #1{\unskip}     \fi
\ifx \showLCCN     \undefined \def \showLCCN      #1{\unskip}     \fi
\ifx \shownote     \undefined \def \shownote      #1{#1}          \fi
\ifx \showarticletitle \undefined \def \showarticletitle #1{#1}   \fi
\ifx \showURL      \undefined \def \showURL       #1{#1}          \fi
\providecommand\bibfield[2]{#2}
\providecommand\bibinfo[2]{#2}
\providecommand\natexlab[1]{#1}
\providecommand\showeprint[2][]{arXiv:#2}

\bibitem[\protect\citeauthoryear{??}{int}{}]%
        {intelberkeley}
\showarticletitle{Intel Lab Data}. In \bibinfo{booktitle}{{\em URL:
  http://db.csail.mit.edu/labdata/labdata.html}}.
\newblock


\bibitem[\protect\citeauthoryear{Acharya, Gibbons, Poosala, and
  Ramaswamy}{Acharya et~al\mbox{.}}{1999}]%
        {acharya1999join}
\bibfield{author}{\bibinfo{person}{Swarup Acharya}, \bibinfo{person}{Phillip~B
  Gibbons}, \bibinfo{person}{Viswanath Poosala}, {and} \bibinfo{person}{Sridhar
  Ramaswamy}.} \bibinfo{year}{1999}\natexlab{}.
\newblock \showarticletitle{{Join Synopses for Approximate Query Answering}}.
  In \bibinfo{booktitle}{{\em SIGMOD}}.
\newblock


\bibitem[\protect\citeauthoryear{Agarwal, Mozafari, et~al\mbox{.}}{Agarwal
  et~al\mbox{.}}{2013}]%
        {agarwal2013blinkdb}
\bibfield{author}{\bibinfo{person}{Sameer Agarwal}, \bibinfo{person}{Barzan
  Mozafari}, {and} \bibinfo{person}{others}.} \bibinfo{year}{2013}\natexlab{}.
\newblock \showarticletitle{{BlinkDB: Queries With Bounded Errors and Bounded
  Response Times on Very Large Data}}. In \bibinfo{booktitle}{{\em EuroSys}}.
\newblock


\bibitem[\protect\citeauthoryear{Al-Kateb, Lee, and Wang}{Al-Kateb
  et~al\mbox{.}}{2007}]%
        {al2007adaptive}
\bibfield{author}{\bibinfo{person}{Mohammed Al-Kateb},
  \bibinfo{person}{Byung~Suk Lee}, {and} \bibinfo{person}{X~Sean Wang}.}
  \bibinfo{year}{2007}\natexlab{}.
\newblock \showarticletitle{{Adaptive-size Reservoir Sampling Over Data
  Streams}}. In \bibinfo{booktitle}{{\em SSDBM}}.
\newblock


\bibitem[\protect\citeauthoryear{Blanas, Li, and Patel}{Blanas
  et~al\mbox{.}}{2011}]%
        {blanas2011design}
\bibfield{author}{\bibinfo{person}{Spyros Blanas}, \bibinfo{person}{Yinan Li},
  {and} \bibinfo{person}{Jignesh~M Patel}.} \bibinfo{year}{2011}\natexlab{}.
\newblock \showarticletitle{Design and evaluation of main memory hash join
  algorithms for multi-core CPUs}. In \bibinfo{booktitle}{{\em SIGMOD}}.
\newblock


\bibitem[\protect\citeauthoryear{Blanas, Patel, Ercegovac, and Rao}{Blanas
  et~al\mbox{.}}{2010}]%
        {blanas2010comparison}
\bibfield{author}{\bibinfo{person}{Spyros Blanas}, \bibinfo{person}{Jignesh~M
  Patel}, \bibinfo{person}{Vuk Ercegovac}, {and} \bibinfo{person}{Jun Rao}.}
  \bibinfo{year}{2010}\natexlab{}.
\newblock \showarticletitle{A comparison of join algorithms for log processing
  in mapreduce}. In \bibinfo{booktitle}{{\em SIGMOD}}.
\newblock


\bibitem[\protect\citeauthoryear{Chaudhuri, Motwani, and Narasayya}{Chaudhuri
  et~al\mbox{.}}{1999}]%
        {chaudhuri1999random}
\bibfield{author}{\bibinfo{person}{Surajit Chaudhuri}, \bibinfo{person}{Rajeev
  Motwani}, {and} \bibinfo{person}{Vivek Narasayya}.}
  \bibinfo{year}{1999}\natexlab{}.
\newblock \showarticletitle{{On Random Sampling over Joins}}. In
  \bibinfo{booktitle}{{\em SIGMOD}}.
\newblock


\bibitem[\protect\citeauthoryear{Cho, Myers, and Leskovec}{Cho
  et~al\mbox{.}}{2011}]%
        {cho2011friendship}
\bibfield{author}{\bibinfo{person}{Eunjoon Cho}, \bibinfo{person}{Seth~A
  Myers}, {and} \bibinfo{person}{Jure Leskovec}.}
  \bibinfo{year}{2011}\natexlab{}.
\newblock \showarticletitle{{Friendship and Mobility: User Movement in
  Location-based Social Networks}}. In \bibinfo{booktitle}{{\em SIGKDD}}.
\newblock


\bibitem[\protect\citeauthoryear{Cochran}{Cochran}{2007}]%
        {cochran2007sampling}
\bibfield{author}{\bibinfo{person}{William~G Cochran}.}
  \bibinfo{year}{2007}\natexlab{}.
\newblock \showarticletitle{{Sampling Techniques}}. \bibinfo{publisher}{John
  Wiley \& Sons}.
\newblock


\bibitem[\protect\citeauthoryear{D{\i}az-Garc{\i}a and
  Garay-T{\'a}pia}{D{\i}az-Garc{\i}a and Garay-T{\'a}pia}{2005}]%
        {diazoptimum}
\bibfield{author}{\bibinfo{person}{Jos{\'e}~A D{\i}az-Garc{\i}a} {and}
  \bibinfo{person}{Ma~Magdalena Garay-T{\'a}pia}.}
  \bibinfo{year}{2005}\natexlab{}.
\newblock \showarticletitle{{Optimum Allocation in Stratified Surveys}}. In
  \bibinfo{booktitle}{{\em Comunicaciones del Cimat}}.
\newblock


\bibitem[\protect\citeauthoryear{Dobra, Jermaine, Rusu, and Xu}{Dobra
  et~al\mbox{.}}{2009}]%
        {dobra2009turbo}
\bibfield{author}{\bibinfo{person}{Alin Dobra}, \bibinfo{person}{Chris
  Jermaine}, \bibinfo{person}{Florin Rusu}, {and} \bibinfo{person}{Fei Xu}.}
  \bibinfo{year}{2009}\natexlab{}.
\newblock \showarticletitle{{Turbo-Charging Estimate Convergence in {DBO}}}. In
  \bibinfo{booktitle}{{\em VLDB}}.
\newblock


\bibitem[\protect\citeauthoryear{Estan and Naughton}{Estan and
  Naughton}{2006}]%
        {estan2006end}
\bibfield{author}{\bibinfo{person}{Cristian Estan} {and}
  \bibinfo{person}{Jeffrey~F Naughton}.} \bibinfo{year}{2006}\natexlab{}.
\newblock \showarticletitle{{End-Biased Samples for Join Cardinality
  Estimation}}. In \bibinfo{booktitle}{{\em ICDE}}.
\newblock


\bibitem[\protect\citeauthoryear{Ganguly, Gibbons, Matias, and
  Silberschatz}{Ganguly et~al\mbox{.}}{1996}]%
        {ganguly1996bifocal}
\bibfield{author}{\bibinfo{person}{Sumit Ganguly}, \bibinfo{person}{Phillip~B
  Gibbons}, \bibinfo{person}{Yossi Matias}, {and} \bibinfo{person}{Avi
  Silberschatz}.} \bibinfo{year}{1996}\natexlab{}.
\newblock \showarticletitle{{Bifocal Sampling for Skew-resistant Join Size
  Estimation}}. In \bibinfo{booktitle}{{\em SIGMOD}}.
\newblock


\bibitem[\protect\citeauthoryear{Gemulla, R{\"o}sch, and Lehner}{Gemulla
  et~al\mbox{.}}{2008}]%
        {gemulla2008linked}
\bibfield{author}{\bibinfo{person}{Rainer Gemulla}, \bibinfo{person}{Philipp
  R{\"o}sch}, {and} \bibinfo{person}{Wolfgang Lehner}.}
  \bibinfo{year}{2008}\natexlab{}.
\newblock \showarticletitle{Linked bernoulli synopses: Sampling along foreign
  keys}. In \bibinfo{booktitle}{{\em SSDBM}}.
\newblock


\bibitem[\protect\citeauthoryear{Haas et~al\mbox{.}}{Haas
  et~al\mbox{.}}{1999}]%
        {haas1999ripple}
\bibfield{author}{\bibinfo{person}{Peter~J Haas} {and}
  \bibinfo{person}{others}.} \bibinfo{year}{1999}\natexlab{}.
\newblock \showarticletitle{{Ripple Joins for Online Aggregation}}. In
  \bibinfo{booktitle}{{\em SIGMOD}}.
\newblock


\bibitem[\protect\citeauthoryear{Haas and Haas}{Haas and Haas}{1996}]%
        {haas1996hoeffding}
\bibfield{author}{\bibinfo{person}{Peter~J Haas} {and} \bibinfo{person}{Peter~J
  Haas}.} \bibinfo{year}{1996}\natexlab{}.
\newblock \showarticletitle{{Hoeffding Inequalities for Join-Selectivity
  Estimation and Online Aggregation}}. In \bibinfo{booktitle}{{\em Research
  Report}}. \bibinfo{publisher}{IBM}.
\newblock


\bibitem[\protect\citeauthoryear{Haas, Naughton, Seshadri, et~al\mbox{.}}{Haas
  et~al\mbox{.}}{1996}]%
        {haas1996selectivity}
\bibfield{author}{\bibinfo{person}{Peter~J Haas}, \bibinfo{person}{Jeffrey~F
  Naughton}, \bibinfo{person}{S Seshadri}, {and} \bibinfo{person}{others}.}
  \bibinfo{year}{1996}\natexlab{}.
\newblock \showarticletitle{{Selectivity and Cost Estimation for Joins Based on
  Random Sampling}}. In \bibinfo{booktitle}{{\em JCSS}}.
\newblock


\bibitem[\protect\citeauthoryear{Habich, Lehner, and Hinneburg}{Habich
  et~al\mbox{.}}{2005}]%
        {habich2005optimizing}
\bibfield{author}{\bibinfo{person}{Dirk Habich}, \bibinfo{person}{Wolfgang
  Lehner}, {and} \bibinfo{person}{Alexander Hinneburg}.}
  \bibinfo{year}{2005}\natexlab{}.
\newblock \showarticletitle{Optimizing Multiple Top-K Queries over Joins.}. In
  \bibinfo{booktitle}{{\em SSDBM}}.
\newblock


\bibitem[\protect\citeauthoryear{He, Yang, Fang, Lu, Govindaraju, Luo, and
  Sander}{He et~al\mbox{.}}{2008}]%
        {he2008relational}
\bibfield{author}{\bibinfo{person}{Bingsheng He}, \bibinfo{person}{Ke Yang},
  \bibinfo{person}{Rui Fang}, \bibinfo{person}{Mian Lu}, \bibinfo{person}{Naga
  Govindaraju}, \bibinfo{person}{Qiong Luo}, {and} \bibinfo{person}{Pedro
  Sander}.} \bibinfo{year}{2008}\natexlab{}.
\newblock \showarticletitle{Relational joins on graphics processors}. In
  \bibinfo{booktitle}{{\em SIGMOD}}.
\newblock


\bibitem[\protect\citeauthoryear{Hellerstein et~al\mbox{.}}{Hellerstein
  et~al\mbox{.}}{1997}]%
        {hellerstein1997online}
\bibfield{author}{\bibinfo{person}{Joseph~M Hellerstein} {and}
  \bibinfo{person}{others}.} \bibinfo{year}{1997}\natexlab{}.
\newblock \showarticletitle{{Online Aggregation}}. In \bibinfo{booktitle}{{\em
  SIGMOD}}.
\newblock


\bibitem[\protect\citeauthoryear{Jermaine et~al\mbox{.}}{Jermaine
  et~al\mbox{.}}{2006}]%
        {jermaine2006sort}
\bibfield{author}{\bibinfo{person}{Christopher Jermaine} {and}
  \bibinfo{person}{others}.} \bibinfo{year}{2006}\natexlab{}.
\newblock \showarticletitle{{The Sort-Merge-Shrink Join}}. In
  \bibinfo{booktitle}{{\em TODS}}.
\newblock


\bibitem[\protect\citeauthoryear{Jermaine, Arumugam, Pol, and Dobra}{Jermaine
  et~al\mbox{.}}{2008}]%
        {jermaine2008scalable}
\bibfield{author}{\bibinfo{person}{Chris Jermaine},
  \bibinfo{person}{Subramanian Arumugam}, \bibinfo{person}{Abhijit Pol}, {and}
  \bibinfo{person}{Alin Dobra}.} \bibinfo{year}{2008}\natexlab{}.
\newblock \showarticletitle{{Scalable Approximate Query Processing with the
  {DBO} Engine}}. In \bibinfo{booktitle}{{\em TODS}}.
\newblock


\bibitem[\protect\citeauthoryear{Jermaine, Dobra, Arumugam, Joshi, and
  Pol}{Jermaine et~al\mbox{.}}{2005}]%
        {jermaine2005disk}
\bibfield{author}{\bibinfo{person}{Christopher Jermaine}, \bibinfo{person}{Alin
  Dobra}, \bibinfo{person}{Subramanian Arumugam}, \bibinfo{person}{Shantanu
  Joshi}, {and} \bibinfo{person}{Abhijit Pol}.}
  \bibinfo{year}{2005}\natexlab{}.
\newblock \showarticletitle{{A Disk-based Join with Probabilistic Guarantees}}.
  In \bibinfo{booktitle}{{\em SIGMOD}}.
\newblock


\bibitem[\protect\citeauthoryear{Kadane}{Kadane}{2005}]%
        {kadane2005optimal}
\bibfield{author}{\bibinfo{person}{Joseph~B Kadane}.}
  \bibinfo{year}{2005}\natexlab{}.
\newblock \showarticletitle{{Optimal Dynamic Sample Allocation among Strata}}.
  In \bibinfo{booktitle}{{\em Journal of official Statistics}}.
\newblock


\bibitem[\protect\citeauthoryear{Kandula, Shanbhag, Vitorovic,
  et~al\mbox{.}}{Kandula et~al\mbox{.}}{2016}]%
        {kandula2016quickr}
\bibfield{author}{\bibinfo{person}{Srikanth Kandula}, \bibinfo{person}{Anil
  Shanbhag}, \bibinfo{person}{Aleksandar Vitorovic}, {and}
  \bibinfo{person}{others}.} \bibinfo{year}{2016}\natexlab{}.
\newblock \showarticletitle{{Quickr: Lazily Approximating Complex AdHoc Queries
  in BigData Clusters}}. In \bibinfo{booktitle}{{\em SIGMOD}}.
\newblock


\bibitem[\protect\citeauthoryear{Kent}{Kent}{1994}]%
        {kent1994sloan}
\bibfield{author}{\bibinfo{person}{Stephen~M Kent}.}
  \bibinfo{year}{1994}\natexlab{}.
\newblock \showarticletitle{{Sloan Digital Sky Survey}}. In
  \bibinfo{booktitle}{{\em Science with Astronomical Near-infrared Sky
  Surveys}}.
\newblock


\bibitem[\protect\citeauthoryear{Kitikidou}{Kitikidou}{2012}]%
        {kitikidou2012optimizing}
\bibfield{author}{\bibinfo{person}{Kyriaki Kitikidou}.}
  \bibinfo{year}{2012}\natexlab{}.
\newblock \showarticletitle{{Optimizing Forest Sampling by using Lagrange
  Multipliers}}. In \bibinfo{booktitle}{{\em AJOR}}.
\newblock


\bibitem[\protect\citeauthoryear{Li, Wu, Yi, and Zhao}{Li
  et~al\mbox{.}}{2016}]%
        {li2016wander}
\bibfield{author}{\bibinfo{person}{Feifei Li}, \bibinfo{person}{Bin Wu},
  \bibinfo{person}{Ke Yi}, {and} \bibinfo{person}{Zhuoyue Zhao}.}
  \bibinfo{year}{2016}\natexlab{}.
\newblock \showarticletitle{{Wander Join: Online Aggregation via Random
  Walks}}. In \bibinfo{booktitle}{{\em SIGMOD}}.
\newblock


\bibitem[\protect\citeauthoryear{Lohr}{Lohr}{2010}]%
        {lohr2010sampling}
\bibfield{author}{\bibinfo{person}{Sharon~L Lohr}.}
  \bibinfo{year}{2010}\natexlab{}.
\newblock \showarticletitle{{Sampling: Design and Analysis}}.
\newblock


\bibitem[\protect\citeauthoryear{Luo et~al\mbox{.}}{Luo et~al\mbox{.}}{2002}]%
        {luo2002scalable}
\bibfield{author}{\bibinfo{person}{Gang Luo} {and} \bibinfo{person}{others}.}
  \bibinfo{year}{2002}\natexlab{}.
\newblock \showarticletitle{{A Scalable Hash Ripple Join Algorithm}}. In
  \bibinfo{booktitle}{{\em SIGMOD}}.
\newblock


\bibitem[\protect\citeauthoryear{Mokbel, Lu, and Aref}{Mokbel
  et~al\mbox{.}}{2004}]%
        {mokbel2004hash}
\bibfield{author}{\bibinfo{person}{Mohamed~F Mokbel}, \bibinfo{person}{Ming
  Lu}, {and} \bibinfo{person}{Walid~G Aref}.} \bibinfo{year}{2004}\natexlab{}.
\newblock \showarticletitle{Hash-merge join: A non-blocking join algorithm for
  producing fast and early join results}. In \bibinfo{booktitle}{{\em ICDE}}.
\newblock


\bibitem[\protect\citeauthoryear{Nirkhiwale, Dobra, and Jermaine}{Nirkhiwale
  et~al\mbox{.}}{2013}]%
        {nirkhiwale2013sampling}
\bibfield{author}{\bibinfo{person}{Supriya Nirkhiwale}, \bibinfo{person}{Alin
  Dobra}, {and} \bibinfo{person}{Christopher Jermaine}.}
  \bibinfo{year}{2013}\natexlab{}.
\newblock \showarticletitle{{A Sampling Algebra for Aggregate Estimation}}. In
  \bibinfo{booktitle}{{\em VLDB}}.
\newblock


\bibitem[\protect\citeauthoryear{Olken}{Olken}{1993}]%
        {olken1993random}
\bibfield{author}{\bibinfo{person}{Frank Olken}.}
  \bibinfo{year}{1993}\natexlab{}.
\newblock \showarticletitle{{Random Sampling from Databases}}.
\newblock


\bibitem[\protect\citeauthoryear{Qin and Rusu}{Qin and Rusu}{2014}]%
        {qin2014pf}
\bibfield{author}{\bibinfo{person}{Chengjie Qin} {and} \bibinfo{person}{Florin
  Rusu}.} \bibinfo{year}{2014}\natexlab{}.
\newblock \showarticletitle{{PF-OLA: A High-performance Framework for Parallel
  Online Aggregation}}. In \bibinfo{booktitle}{{\em Distributed and Parallel
  Databases}}.
\newblock


\bibitem[\protect\citeauthoryear{Ray, Simion, Brown, and Johnson}{Ray
  et~al\mbox{.}}{2014}]%
        {ray2014skew}
\bibfield{author}{\bibinfo{person}{Suprio Ray}, \bibinfo{person}{Bogdan
  Simion}, \bibinfo{person}{Angela~Demke Brown}, {and} \bibinfo{person}{Ryan
  Johnson}.} \bibinfo{year}{2014}\natexlab{}.
\newblock \showarticletitle{Skew-resistant parallel in-memory spatial join}. In
  \bibinfo{booktitle}{{\em SSDBM}}.
\newblock


\bibitem[\protect\citeauthoryear{Shin, Moon, and Lee}{Shin
  et~al\mbox{.}}{2000}]%
        {shin2000adaptive}
\bibfield{author}{\bibinfo{person}{Hyoseop Shin}, \bibinfo{person}{Bongki
  Moon}, {and} \bibinfo{person}{Sukho Lee}.} \bibinfo{year}{2000}\natexlab{}.
\newblock \showarticletitle{Adaptive multi-stage distance join processing}. In
  \bibinfo{booktitle}{{\em SIGMOD}}.
\newblock


\bibitem[\protect\citeauthoryear{Sidirourgos, Kersten, and Boncz}{Sidirourgos
  et~al\mbox{.}}{2011}]%
        {sidirourgos2011sciborq}
\bibfield{author}{\bibinfo{person}{Lefteris Sidirourgos},
  \bibinfo{person}{Martin~L Kersten}, {and} \bibinfo{person}{Peter~A Boncz}.}
  \bibinfo{year}{2011}\natexlab{}.
\newblock \showarticletitle{{SciBORQ: Scientific Data Management with Bounds On
  Runtime and Quality}}. In \bibinfo{booktitle}{{\em CIDR}}.
\newblock


\bibitem[\protect\citeauthoryear{Spiegel and Polyzotis}{Spiegel and
  Polyzotis}{2009}]%
        {spiegel2009tug}
\bibfield{author}{\bibinfo{person}{Joshua Spiegel} {and}
  \bibinfo{person}{Neoklis Polyzotis}.} \bibinfo{year}{2009}\natexlab{}.
\newblock \showarticletitle{{TuG Synopses for Approximate Query Answering}}. In
  \bibinfo{booktitle}{{\em TODS}}.
\newblock


\bibitem[\protect\citeauthoryear{Sukhatme}{Sukhatme}{1957}]%
        {sukhatme1957sampling}
\bibfield{author}{\bibinfo{person}{Pandurang Sukhatme}.}
  \bibinfo{year}{1957}\natexlab{}.
\newblock \bibinfo{booktitle}{{\em {Sampling Theory of Surveys with
  Applications}}}.
\newblock \bibinfo{publisher}{ISAS}.
\newblock


\bibitem[\protect\citeauthoryear{Vengerov, Menck, Zait, and
  Chakkappen}{Vengerov et~al\mbox{.}}{2015}]%
        {vengerov2015join}
\bibfield{author}{\bibinfo{person}{David Vengerov},
  \bibinfo{person}{Andre~Cavalheiro Menck}, \bibinfo{person}{Mohamed Zait},
  {and} \bibinfo{person}{Sunil~P Chakkappen}.} \bibinfo{year}{2015}\natexlab{}.
\newblock \showarticletitle{{Join Size Estimation Subject to Filter
  Conditions}}. In \bibinfo{booktitle}{{\em VLDB}}.
\newblock


\bibitem[\protect\citeauthoryear{Vitter}{Vitter}{1985}]%
        {vitter1985random}
\bibfield{author}{\bibinfo{person}{Jeffrey~S Vitter}.}
  \bibinfo{year}{1985}\natexlab{}.
\newblock \showarticletitle{Random sampling with a reservoir}. In
  \bibinfo{booktitle}{{\em TOMS}}.
\newblock


\bibitem[\protect\citeauthoryear{Wah and Wu}{Wah and Wu}{1999}]%
        {wah1999theory}
\bibfield{author}{\bibinfo{person}{Benjamin~W Wah} {and} \bibinfo{person}{Zhe
  Wu}.} \bibinfo{year}{1999}\natexlab{}.
\newblock \showarticletitle{{The Theory of Discrete Lagrange Multipliers for
  Nonlinear Discrete Optimization}}. In \bibinfo{booktitle}{{\em CP}}.
  Springer.
\newblock


\bibitem[\protect\citeauthoryear{Wickham}{Wickham}{2011}]%
        {wickham2011asa}
\bibfield{author}{\bibinfo{person}{Hadley Wickham}.}
  \bibinfo{year}{2011}\natexlab{}.
\newblock \showarticletitle{ASA 2009 Data Expo, Flights Dataset}.
\newblock


\bibitem[\protect\citeauthoryear{Yu, Hou, Luo, Che, and Zhu}{Yu
  et~al\mbox{.}}{2013}]%
        {yu2013cs2}
\bibfield{author}{\bibinfo{person}{Feng Yu}, \bibinfo{person}{Wen-Chi Hou},
  \bibinfo{person}{Cheng Luo}, \bibinfo{person}{Dunren Che}, {and}
  \bibinfo{person}{Mengxia Zhu}.} \bibinfo{year}{2013}\natexlab{}.
\newblock \showarticletitle{{CS2: A New Database Synopsis for Query
  Estimation}}. In \bibinfo{booktitle}{{\em SIGMOD}}.
\newblock


\bibitem[\protect\citeauthoryear{Zhang, Chen, and Wang}{Zhang
  et~al\mbox{.}}{2012}]%
        {zhang2012towards}
\bibfield{author}{\bibinfo{person}{Xiaofei Zhang}, \bibinfo{person}{Lei Chen},
  {and} \bibinfo{person}{Min Wang}.} \bibinfo{year}{2012}\natexlab{}.
\newblock \showarticletitle{Towards efficient join processing over large RDF
  graph using mapreduce}. In \bibinfo{booktitle}{{\em SSDBM}}.
\newblock


\bibitem[\protect\citeauthoryear{Zhao, Rusu, Dong, et~al\mbox{.}}{Zhao
  et~al\mbox{.}}{2016}]%
        {zhao2016similarity}
\bibfield{author}{\bibinfo{person}{Weijie Zhao}, \bibinfo{person}{Florin Rusu},
  \bibinfo{person}{Bin Dong}, {and} \bibinfo{person}{others}.}
  \bibinfo{year}{2016}\natexlab{}.
\newblock \showarticletitle{Similarity Join over Array Data}. In
  \bibinfo{booktitle}{{\em Proceedings of the 2016 International Conference on
  Management of Data}}.
\newblock


\end{thebibliography}
 
\end{document}